\documentclass[twocolumn,aps]{emulateapj09}

\newcommand{\Hu}{{\cal H}}

\newcommand{\0}{{^{(0)}}}

\newcommand{\jcap}{Journal of Cosmology and Astroparticle Physics}

\usepackage{graphicx}
\usepackage{epsfig}
\usepackage{amsmath}
\usepackage{color}
\usepackage{hyperref}
\usepackage{natbib}

\shorttitle{ Gravity Parameter Space}
\shortauthors{Baker et al.} 

\begin{document}

\submitted{ApJ 802, 63; published 23rd March 2015.}

\title{Linking Tests of Gravity On All Scales: from the Strong-Field Regime to Cosmology}

\author{Tessa Baker\altaffilmark{1}, Dimitrios Psaltis\altaffilmark{2} and Constantinos Skordis\altaffilmark{3,4}}

\altaffiltext{1}{Astrophysics, Denys Wilkinson Building, Keble Road, University of Oxford, Oxford, OX1 3RH, UK.}

\altaffiltext{2}{Astronomy Department, University of Arizona, 933 N. Cherry Ave., Tucson, AZ 85721, USA.}

\altaffiltext{3}{School of Physics and Astronomy, University of Nottingham, Nottingham NG7 2RD, UK.}

\altaffiltext{4}{Department of Physics, University of Cyprus, 1 University Avenue, Nicosia 2109, Cyprus.}

\email{tessa.baker@astro.ox.ac.uk}
\email{dpsaltis@email.arizona.edu}
\email{skordis@ucy.ac.cy}

\begin{abstract}

The current effort to test General Relativity employs multiple disparate formalisms for different observables, obscuring the relations between laboratory, astrophysical and cosmological constraints. To remedy this situation, we develop a parameter space for comparing tests of gravity on all scales in the universe. In particular, we present new methods for linking cosmological large-scale structure, the Cosmic Microwave Background and gravitational waves with classic PPN tests of gravity. Diagrams of this gravitational parameter space reveal a noticeable untested regime. The untested window, which separates small-scale systems from the troubled cosmological regime, could potentially hide the onset of corrections to General Relativity. %
\end{abstract}


\section{INTRODUCTION}
\label{section:intro}
General Relativity (GR) is unique amongst our fundamental theories of physics, in that it does not fail below the Planck scale -- at least to our present knowledge. That is, we are not aware of any fundamental scale (other than M$_{\mathrm{Pl}}$) associated with the onset of new gravitational physics. This is in contrast to, say, the separate theories of electromagnetism and the weak nuclear force (which fuse at the electroweak scale of $\sim 200$~GeV)\footnote{Furthermore, we know that, in a different limit,  electromagnetism gives way to quantum electrodynamics.}, and most probably the separate electroweak and strong nuclear forces, which are believed to unify at the GUT scale of $\sim 10^{16}$~GeV. 

The apparent infallibility of GR could, of course, indicate that it is a complete theory, valid at all energy scales up to the Planck scale. However, the tempting possibility of explaining the accelerating expansion rate of the universe through modifications to GR (see \citealt{Clifton2012} for a review) suggests that it is worth considering otherwise. The keystone to testing this hypothesis is the identification of a scale at which GR acquires correction terms, and an alternative theory of gravity takes over.

However, there is no \textit{a priori} reason why this transition from GR to modified gravity should happen at a particular \textit{energy} scale. Other physical parameters could instead provide the trigger. For example, in $f\left(R\right)$ gravity, modifications to GR are controlled by the scalar curvature of a spacetime \citep{Sotiriou2010}. Other authors have posited the existence of a fundamental acceleration scale \citep{Bekenstein2004}. Further possibilities include density-dependent \citep{Khoury2004} or lengthscale-dependent behaviour (see \citealt{BakerQS} for further discussion).

One might argue that this distinction has no meaning -- densities, distances, curvatures, etc. can all be expressed as equivalent energies. We would counter that this distinction \textit{is} important, at least for the prosaic task of compiling current experimental bounds on gravity. For example, the surface of the Sun and cosmological-scale density perturbations both probe gravitational potential wells of order \mbox{$\Phi\sim10^{-5}$}. If local gravitational potential is the fundamental quantity controlling modifications to GR, then the Cosmic Microwave Background (CMB) and solar physics both probe the same regime of gravity. In contrast, if acceleration is the fundamental controlling quantity, then the CMB and the solar surface can be used to test gravity in two very different regimes.

This issue is a severe source of difficulty for the research community interested in testing alternative theories of gravity. There is a healthy diversification of work, ranging from `small-scale' tests of gravity based on laboratory and lunar experiments, Solar System satellites, stars and compact objects to large-scale probes such as galaxy correlation functions, galaxy weak lensing, the CMB, CMB lensing and galaxy clusters (and passing the intermediate regime of galaxy rotation curves and strong lensing along the way). It is very likely that in the near future direct detection of gravitational waves will join this list.

The downside to this proliferation is that it is often difficult to connect the constraints on GR obtained from experiments on different scales. It is not obvious that a gravitational theory which is well-behaved on cosmological scales permits the existence of viable stellar solutions (see, for example, \citealt{BarausseSM2008}). Likewise, some theories that give testable predictions for, say, orbits around black holes may be unable to produce cosmic acceleration\footnote{To rectify the situation a cosmological constant term is often added to the theory, either explicitly or in a `backdoor' manner. Occam's razor complains loudly in such cases.}. This state of affairs is further hindered by the postulate of screening mechanisms \citep{Vainshtein1972,Khoury2004,Braxdilaton}, phenomena which act to suppress any non-GR behaviour in particular environments\footnote{It is by no means guaranteed that a screening mechanism can be embedded into all theories of gravity. Given that only a handful of explicit screening mechanisms are known to date, it is more conservative to assume that the majority of gravity theories do \textit{not} screen.}.

The purpose of this paper is to introduce a well-defined, quantitative procedure for comparing the environments probed by different tests of gravity (see \citealt{Psaltis2008} for earlier ideas along these lines). Placing all systems on a common set of axes should facilitate discussion between different sectors of the gravitational physics community. Furthermore, making plain the remit of existing constraints will allow us to sensibly ask the question: is there still `room' for departures from GR in the present state of affairs? Are there untested gravitational environments that might provide the most fruitful directions for future research?

{We stress from the outset that this paper does not address issues of experimental difficulty involved in performing a precision test of gravity. In many of the situations we will discuss, astrophysical systematics dominate the relativistic effects by orders of magnitude. However, our optimistic attitude is motivated by recent examples in which such systematics have been successfully modelled and subtracted. For example, in a test of gravity using radio links with the Cassini spacecraft, successful removal of dominating noise from the solar coronal plasma resulted in systematic errors four orders of magnitude smaller than the relativistic signal \citep{Bertotti2003}. Similarly, the incredible precision of current pulsar constraints is obtained using detailed modelling of a series of gravitational interaction terms and orbital delays. On the cosmological front, N-body simulations are used to model nonlinear and baryonic effects. There are clear goals set for the improvements needed to deal with data from the next generation of cosmological experiments (approximately a $\sim 1\%$ accuracy on the matter power spectrum, \citealt{Semboloni2013}). 

In \S\ref{section:quant} we explain our choice of axes for a gravitational parameter space, and how both astrophysical and cosmological systems can be mapped onto them. \S\ref{section:plot} is devoted to the understanding of this plot. In \S\ref{section:experiments} we provide a visualization of \textit{experimental} constraints, which indicates where the under-tested regimes of gravity lie. \S\ref{section:conclusions} is devoted to a discussion of our results.

In this paper we will work in conformal time $\eta$, denoting derivatives with respect to $\eta$ by a dot. The conformal Hubble factor is \mbox{$\Hu\equiv\dot{a}/a$}. Fractional energy densities such as $\Omega_M(\eta)$ denote time-dependent quantities; present-day values are indicated by a subscript zero, e.g. $\Omega_{M0}$. The metric signature used is \mbox{$\{-,+,+,+\}$}. Some extended calculations are sequestered in the Appendix.\newline

\section{QUANTIFYING GRAVITATIONAL FIELDS}
\label{section:quant}

\subsection{Categories of Systems}
\label{sub:systems}
The gravitational systems considered in this paper fall into three categories: laboratory, astrophysical and cosmological. Most of our discussion will focus on the latter two categories.

The astrophysical systems are nearly all spherically symmetric, and many can be approximated by a test particle in orbit around a central mass, e.g. a planet orbiting a star, a star orbiting close to a supermassive black hole, etc. Observations of the test particle's motion are considered as a probe of the gravitational field of the larger body.

Cosmological systems, e.g. the CMB, must instead be treated as power spectra. These require more careful handling; a gravitational field must be assigned to each wavenumber $k$ or angular mode $\ell$ in the power spectrum. We need to define quantifiers analogous to those applied to astrophysical systems, so that comparisons between the two categories are possible.

Below we set out the system we will use to assess gravitational field strengths for the astrophysical and cosmological categories. In \S\ref{sub:laboratory} we will explain how equivalent parameters are assigned to two specific laboratory tests of gravity.

\subsection{Gravitational Quantifiers} 
\label{sub:3numbers}

In GR, three tensors make up the description of spacetime that enters the Einstein equations: the metric, the Riemann curvature tensor, and the Ricci tensor\footnote{We regard the stress-energy tensor of matter as \textit{sourcing} the curvature of spacetime, not as part of its description.}.
We can characterize a gravitational field by assessing how it is distributed between these three tensors. A loose physical interpretation runs thus: the metric describes the curvature of the spacetime at a point; the Ricci tensor describes how much of that curvature can be attributed to the local mass at that point (since the Ricci tensor vanishes in vacuum); the Riemann curvature tensor describes the total curvature due to both local masses and the gravitational fields of other masses at a distance. 
  
We wish to construct scalars which quantify the relative importance of each of these three tensors for a given gravitational field. However, the obvious choice for the Ricci tensor -- the canonical Ricci scalar -- vanishes in vacuum and radiation-dominated systems, making it awkward for the purposes of this paper\footnote{For example, $R$ does not distinguish between a particle in orbit around a black hole (a vacuum situation) and the early universe (a radiation-dominated situation), since it is zero in both cases. }$^{,}$\footnote{Note that the other semi-obvious choice, $\left(R_{\alpha\beta}R^{\alpha\beta}\right)^{\frac{1}{2}}$, similarly vanishes.}. Hence we will focus our attention on the remaining two tensors, the metric and the Riemann curvature tensor. 

Let us first consider the example of a test particle situated at a radial distance $r$ from a central object of mass $M$. The deviation of the metric from Minkowski form is characterized by the magnitude of the Newtonian gravitational potential,
\begin{align}
\epsilon\equiv \frac{GM}{rc^2}\;.
\label{eq:potential}
\end{align}
The strongest gravitational fields accessible to an observer correspond to the limit $\epsilon\rightarrow {\cal O}(1)$, when the central object is a black hole and the test particle orbits close to the event horizon. Although equation (\ref{eq:potential}) is a coordinate-dependent statement, it can be linked to a directly observable (and therefore coordinate-independent) quantity, namely the gravitational redshift of emission lines from a star or similar object. Hence equation (\ref{eq:potential}) is a valid parameter for assessing the approximate magnitude of the components of the metric outside a single object in vacuum.

We will measure the approximate magnitude of the Riemann curvature tensor through the Kretschmann scalar $(R^{\alpha\beta\gamma\delta}R_{\alpha\beta\gamma\delta})^{1/2}$. The Kretschmann scalar for the
Schwarzschild metric is
\begin{align}
\label{eq:xi_astro}
\xi=\left(R^{\alpha\beta\gamma\delta}R_{\alpha\beta\gamma\delta}\right)^{1/2}&=
\sqrt{48}\;\frac{GM}{r^3c^2}\;.
\end{align}
The first equality above is coordinate-independent, and serves as our formal definition of $\xi$; the second equality is merely an illustratory example for the choice of standard Schwarzchild coordinates.
The corresponding expression for rotating objects is more complicated \citep{Henry2000}. However, the additional prefactors will make little difference on the axis ranges used in this paper (see Fig.~\ref{fig:parameter_space}), and so will be neglected.

The parameters $\epsilon$ and $\xi$ above define a two-dimensional space on which we can place the gravitational fields probed by different objects, observations and experiments. For simplicity we will informally refer to these parameters as the `curvature' and the `potential' of the spacetime, though this is not strictly accurate in all the contexts we consider. We stress that our parameters $\epsilon$ and $\xi$ depart from the \textit{physical} potential and curvature in some regimes, e.g. the metric for the interior of a star. However, they still serve as useful yardsticks that allow us to compare radically different gravitational environments.

One might question if there is some redundancy or degeneracy in the use of $\epsilon$ and $\xi$. After all, $\xi$ is quantifying the Riemann tensor, which is simply a contraction of the metric; and the metric is quantified by $\epsilon$. However, the Riemann tensor also assesses the derivatives of the metric, and hence contains crucial extra information. As a concrete example, consider two black holes, one of stellar mass ($\sim 10~$M$_\sun$) and one supermassive ($\sim 10^7~$M$_\sun$). Consider the gravitational field on the event horizon of these black holes. In both cases the potential is of order unity, but the curvature of the spacetime is very different. Specifically, the $\xi$-value assigned to the supermassive black hole will be lower than that for the stellar mass black hole, because the radius of curvature of the horizon is larger. Hence our two scalar quantifiers are indeed sensitive to different aspects of a gravitational field.

It should be noted though, that there are some aspects of gravitational fields which are not encapsulated by either of our two parameters. Most notably, our scheme does not distinguish between approximately static and dynamical gravitational fields. Most of the systems we consider fall into the former category, but binary pulsars (\S\ref{sub:stellar}) and gravitational wave experiments (\S\ref{sub:grav_waves}) belong to the latter. This enables them to potentially constrain kinetic terms and spin-2 perturbations of a modified gravitational action, terms which are inaccessible in other experiments.

\subsection{Adaptations for Cosmological Perturbations}
\label{sub:cosmo_params}
Equations (\ref{eq:potential}) and (\ref{eq:xi_astro}) are appropriate for gravitational systems that can be modelled as spherically symmetric. In this subsection we describe how we can associate analogous characteristic scales in gravitational potential and Kretschmann curvature to the cosmological perturbations.

We assume that a spectrum of primordial perturbations is laid down by some mechanism in the early universe. These primordial potentials source the growth of matter overdensities, so that at later epochs the matter power spectrum can be used as a tracer for the power spectrum of potential wells in the universe. In GR, the connection between potential wells and matter overdensities is given by the Poisson equation,
\begin{align}
\label{giPoisson1}
\nabla^2\hat\Phi&=\sum_i 4\pi G_N a^2\rho_i\left[\delta_i+3\frac{\Hu}{k^2}(1+\omega_i)\theta_i\right]\\
&=\sum_i 4\pi G_N a^2\rho_i\Delta_i\;.
\label{giPoisson2}
\end{align}
The sums in the expressions above are taken over all cosmologically-relevant fluids (baryons, cold dark matter, etc.). The fractional matter overdensity is \mbox{$\delta\equiv\delta\rho/\rho$}  and the velocity potential $\theta$ is related to the fluid velocity perturbation by $\vec{v}\equiv\vec{\nabla}\theta$. The hat on $\hat\Phi$ indicates that it is a fully gauge-invariant Bardeen potential \citep{Bardeen1980}. 

The second equality above defines the gauge-invariant density perturbation $\Delta$. Note that the second term in equation (\ref{giPoisson1}) causes $\Delta$ to depart from the intuitive matter overdensity $\delta$ on near-horizon scales. However, since only gauge-invariant quantities are observable, we continue to work in terms of $\Delta$. 

We need our cosmological version of $\epsilon$ to represent a statistical average of the range of potentials present in the universe, as a function of redshift. To achieve this, we first consider the time-dependent average:
\begin{align}
\label{epsaverage}
\epsilon(a)&=\sqrt{\langle |\Phi(\vec{x},a) |^2\rangle}\:.
\end{align}
We express the gravitational potential in terms of its Fourier transform, simultaneously introducing a window function in $k$-space, $W(k)$, to account for the fact that real experiments cannot access all wavemodes. This leads to (after evaluating one of the Fourier integrals):
\begin{align}
\epsilon(a)&=\sqrt{\int_0^\infty \frac{d^3k}{(2\pi)^3} \langle|\tilde{\Phi}(\vec{k},a)|^2\rangle \,W(k)}\;.
\label{ark}
\end{align}
Hereafter we will omit tildes from our notation, i.e. we use the same symbol for a variable and its Fourier transform. From the Fourier-space Poisson equation we obtain:
\begin{align}
\label{soj}
\langle\Phi(\vec{k},a)\Phi^*(\vec{k}^\prime,a)\rangle&=\left(\frac{3}{2}\frac{H_0^2\Omega_{M0}}{a}\right)^2 \frac{(2\pi)^3\delta^3(\vec{k}-\vec{k}^\prime)}{|k|^2 |k^{\prime}|^{2}}P_M(k,a)\;,
\end{align}
where $P_M(k,a)$ is the dimensionful power spectrum of the gauge-invariant density perturbation ($\Delta$), such that
\begin{align}
\langle\Delta_M(\vec{k},a)\Delta_M(\vec{k}^\prime,a)\rangle&=(2\pi)^3\delta^3(\vec{k}-\vec{k}^\prime)P_M(k,a)
\end{align}
 and we have used the fact that, for a universe dominated by matter and a cosmological constant,
\begin{align}
8\pi G_N a^2\rho_M=3\Hu^2\Omega_M=3H_0^2\Omega_{M0}/a\;.
\end{align}
Substituting equation (\ref{soj}) into equation (\ref{ark}) results in:
 \begin{align}
\label{realspace_conv1}
\epsilon(a)&=\frac{3}{2}\frac{H_0^2\Omega_{M0}}{a}\left\{\int^\infty_{-\infty}\frac{d\ln k}{2\pi^2}\,\frac{1}{|k|}\,P_M(k,a)W(k)\right\}^{\frac{1}{2}}\;.
\end{align}
For most choices of window function, the above expression will compress an experiment to a single value of $\epsilon$. However, in order to illustrate the full extent of our parameter space that is probed by cosmological surveys, we will choose the simple window function $W(k)=\delta(\ln k-\ln k^\prime)$ (note that this should be substituted in at the stage of equation \ref{ark}). Then the above steps result in a function of both time and scale:
\begin{align}
\label{epsilon_cosmo}
\epsilon^{\mathrm{cosmo}}(k,a)&= \left(\frac{3}{2}\frac{H_0^2\Omega_{M0}}{a}\right)\sqrt{\frac{P_{M}(k,a)}{2\pi^2 |k|}}\;.
\end{align}
This is the expression we will use to determine the x-axis position of linear cosmological perturbations in Fig.~\ref{fig:parameter_space} (see \S\ref{sub:cosmo_lines}). More sophisticated window functions could be considered by those interested in a particular experiment.

We also need to associate a measure of the Kretschmann scalar $(\left[R^{\alpha\beta\gamma\delta}R_{\alpha\beta\gamma\delta}\right]^{1/2})$ to these same cosmological perturbations, as we did for astrophysical systems. However, we now encounter some new subtleties. Firstly, we must use a \textit{gauge-invariant} quantity, as only these are observable (see the discussion following equation \ref{giPoisson2}). 
Secondly, we will separate the Kretchmann scalar into a scale-independent `background' contribution coming from the smooth Friedmann-Robinson-Walker (FRW) metric, and a scale-dependent piece associated to linear cosmological perturbations. That is, we write the Kretschmann scalar as
\begin{align}
\label{kgi}
K(\vec{x},a)&=\0 K(a)+\Xi(\vec{x},a)\;,
\end{align}
where the zeroth-order piece is
 \begin{align}
 \label{K0}
\0 K(a)&=\frac{\sqrt{12}}{a^2}\left(\dot{\Hu}^2+\Hu^4\right)^{\frac{1}{2}}\;,
\end{align}
and $\Xi$ is the perturbative piece of the Kretschmann scalar. In a manner analogous to equation (\ref{epsaverage}), the final quantity we plot is the statistical average: 
\begin{equation}
\label{xidefinitial}
\xi^{\mathrm{cosmo}}(a) = \sqrt{\langle|\hat\Xi(\vec{x},a)|^2\rangle}\;,
\end{equation}
where $\hat\Xi$ is the gauge-invariant counterpart to $\Xi$.

The zeroth-order contribution $\0 K$ dominates over $\xi^{\mathrm{cosmo}}$ considerably. For this reason we will plot them \textit{separately} in the next section (see \S\ref{sub:cosmo_lines}). 
The situation is analogous to that of the CMB temperature anisotropies, for which one generally ignores the temperature monopole of $\sim$2.725K, and instead study the fluctuations around it that are roughly five orders of magnitude smaller\footnote{Let us make it clear that the \textit{physical} Kretschmann curvature that actually exists in the universe is $K_0+\Xi$. However, we will see later that whether we plot the total $K$ or just the perturbative $\Xi$ makes little difference to Fig.~\ref{fig:parameter_space}.}.

For the sake of brevity, we relegate the derivation of the full expression for $\xi^{\mathrm{cosmo}}$ to Appendix~\ref{app:gi_cosmo}. We will show here only the final result, using the same window function as we did to derive equation (\ref{epsilon_cosmo}):
\begin{align}
\xi^{\mathrm{cosmo}}(k,a) = & |A(a)+k^2B(a)| \;  \epsilon^{\mathrm{cosmo}}(k,a)
\label{xi_cosmo}
\end{align}
where $A(a)$ and $B(a)$ are functions of time given in Appendix \ref{app:gi_cosmo}, and $\epsilon^{\mathrm{cosmo}}$ is given by equation (\ref{epsilon_cosmo}). 

Equations (\ref{epsilon_cosmo}) and (\ref{xi_cosmo}) are the expressions we will use to evaluate the gravitational fields probed by cosmological perturbations. In \S\ref{section:experiments} we will discuss how, in practice, these cosmological perturbations are themselves probed using galaxy surveys.

\section{A PARAMETER SPACE FOR GRAVITATIONAL TESTS} 
\label{section:plot}

Fig.~\ref{fig:parameter_space} displays our parameter space
for gravitational systems. The x-axis
of this space corresponds to the potential $\epsilon$
(equation~\ref{eq:potential} or equation~\ref{epsilon_cosmo}, as appropriate) and the y-axis corresponds to the curvature quantity $\xi$ defined in equation (\ref{eq:xi_astro}) or equation (\ref{xi_cosmo}). 

\begin{figure*}[t]
\begin{center}
\hspace{-7mm}
\psfig{file=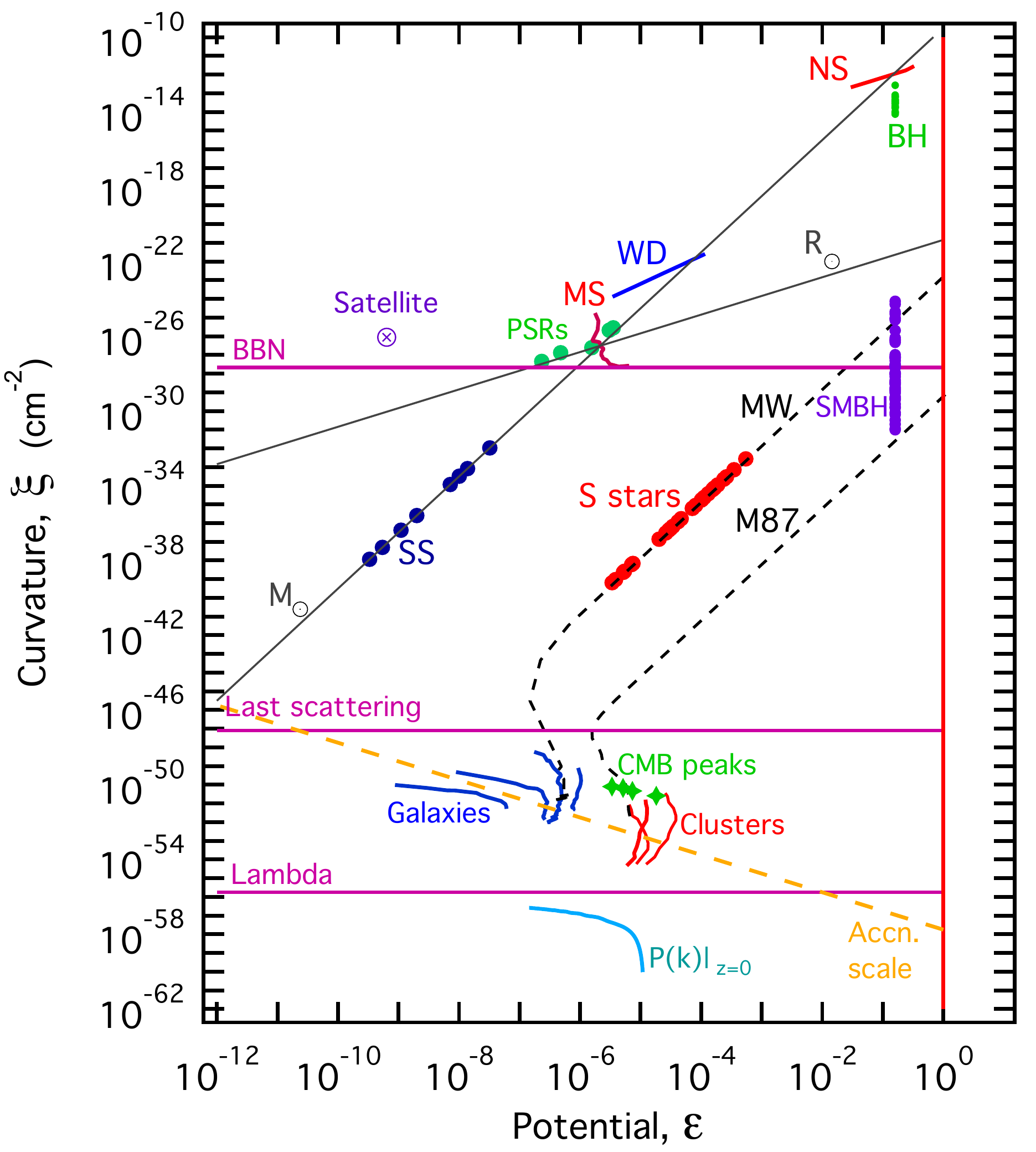,width=15cm}
\caption{A parameter space for gravitational fields, showing the regimes probed by a wide range of astrophysical and cosmological systems. The axes variables are explained in \S\ref{section:quant} and individual curves are detailed in \S\ref{section:plot}. Some of the label abbreviations are: SS = planets of the Solar System, MS = Main Sequence stars, WD = white dwarfs, PSRs = binary pulsars, NS = individual neutron stars, BH = stellar mass black holes, MW = the Milky Way, SMBH = supermassive black holes, BBN = Big Bang Nucleosynthesis.}
\label{fig:parameter_space}
\end{center}
\end{figure*}

\subsection{Understanding the Parameter Space} 

\label{sub:understanding}

The potential in gravitational systems accessible to observers is
bounded by the presence of event horizons at $\epsilon\simeq1$ (we continue to neglect factors of order unity). This limit is shown as a vertical red line in Fig.~\ref{fig:parameter_space}. In principle there is no limit on the maximum curvature accessible by observers, except perhaps the Planck limit, which lies
many orders of magnitude above the bounds of the figure. However, the value of the cosmological constant in our universe defines a minimum curvature; this is indicated by the line labelled `Lambda' in Fig.~\ref{fig:parameter_space}, and described further in \S\ref{sub:cosmo_lines}.

Fig.~\ref{fig:parameter_space} is overlaid with some physically meaningful contours. For example, test particles orbiting the same spherical mass $M$ will lie on a curve of the form (using equations~\ref{eq:potential} and
\ref{eq:xi_astro}):
\begin{align}
\label{constM}
\xi(M)&=\sqrt{48}\,\frac{c^4}{G^2M^2}\epsilon^3\;.
\end{align}
Equation (\ref{constM}) describes a straight line on our logarithmic axes, with an intercept determined by the central
mass $M$. The contour corresponding to $M=1\,M_\odot$ is shown in
Fig.~\ref{fig:parameter_space} (grey). The \textit{innermost} regions of galaxies can also be modelled as test particles orbiting the central black hole, and hence will also be represented by a series of parallel lines (see the description of the Milky Way and M87 curves in \S\ref{sub:gal_cluster} below).

In a similar manner, systems that probe the gravitational fields at a
constant distance $r$ away from central objects with differing masses follow curves of the form
\begin{align}
\xi\left(r\right)&= \frac{\sqrt{48}}{r^2}\epsilon\;.
\end{align}
The contour corresponding to $r=1\,R_\odot$ is shown in
Fig.~\ref{fig:parameter_space} (grey).

In simple Keplerian mechanics, which is adequate for non-relativistic astrophysical systems, the characteristic centripetal acceleration has magnitude $\tilde{a}=GM/r^2$. In terms of our potential and curvature quantifiers, a constant
acceleration therefore corresponds to the curve
\begin{align}
\xi(a) &= \sqrt{48}\left(\frac{\tilde{a}}{c^2}\right)^2 \frac{1}{\epsilon}\;.
\end{align}
This represents a straight line of negative gradient in Fig.~\ref{fig:parameter_space}. It is a well-known but poorly understood phenomenon that the need to introduce dark matter to explain galactic
rotation curves and intra-cluster galaxy velocities coincides
with a fixed acceleration scale of approximately \mbox{$1.2\times
  10^{-10}$ ms$^{-2}$}, which is roughly equivalent to the value of
$c\times H_0$ (\citealt{Bekenstein2004, Famaey2012} and references therein).  The contour corresponding to this particular acceleration scale is shown
in Fig.~\ref{fig:parameter_space} (orange, dashed). Systems below-left of this
acceleration scale cannot be modelled without adding a
contribution to the gravitational field from unseen matter. This region of the parameter space is then problematic\footnote{But not impossibly so, due to the different properties of dark energy and dark matter.} for testing gravity theories, since here there is a degeneracy between two uncertain components of a cosmological model: dark matter and an effective dark energy (which could be due to real fields or corrections to General Relativity).

One final trend is worth noting before we move on to describing specific systems. The gravitational field inside an isothermal sphere with a 
density profile 
\begin{align}
\rho(r)&=\rho_0\left(\frac{r_0}{r}\right)^2
\end{align}
corresponds to a vertical line on the parameter space, since the potential (x-axis) parameter
\begin{align}
\epsilon_{\rm iso}&=\frac{GM(<r)}{rc^2}\\
&=\frac{G\rho_0}{rc^2}
\int_0^r4\pi\left(\frac{r_0}{\tilde{r}}\right)^2\tilde{r}^2d\tilde{r}
\nonumber\\
&=\frac{4\pi G\rho_0r_0^2}{c^2}
\label{virial}
\end{align}
is constant throughout the sphere. This is the reason why the galaxy
cluster profiles and some of the individual galaxy profiles in
Fig.~\ref{fig:parameter_space} are approximately vertical (see also \S\ref{sub:gal_cluster}) -- they represent nearly-virialized systems.

\subsection{Stellar-Scale Objects}
\label{sub:stellar}
We now place individual objects on the parameter space, beginning with some simple test-particle-in-orbit-type systems. When evaluating the potential and curvature probed in these
settings, we use the semi-major axis of the orbit, neglecting any eccentricities as well as gravitational interactions
between multiple orbiting objects.

We also only need to consider the potential well of the dominant mass in the system under consideration. For example, we do not account for the potential well of the Galaxy when considering the potentials probed by
planets orbiting the Sun. This is because only \textit{differences} in
potentials are measurable, and the potential profile of the Galaxy is (to all intents and purposes) constant over the Solar System. The merits and disadvantages of this approach are discussed further in \S\ref{sub:caution}.

Fig.~\ref{fig:parameter_space} and Table~\ref{tab:astro} in Appendix \ref{app:data} show the data for the following astrophysical systems, evaluated using equations (\ref{eq:potential}) and (\ref{eq:xi_astro}): 

\vspace{2mm}
\noindent {\em (i)\/} The marker labelled `satellite' at $\Phi\sim 10^{-9}$ indicates the field characteristics probed by a satellite experiment in a near-Earth orbit, such as Gravity Probe B \citep{GPB2011}.

\vspace{2mm}
\noindent {\em (ii)\/} The points on the Solar System (SS) contour mark
the gravitational fields probed by the local planets, treated as test
particles in orbit around the Sun. As explained in \S\ref{sub:understanding}, the points lie on a straight line because they orbit the same central mass. The non-negligible mass of Jupiter causes a small shift away from this straight line; the shift is not visible on the scale of this plot.

\vspace{2mm}
\noindent {\em (iii)\/} The lines labeled main sequence (MS) and white
dwarfs (WD) mark the gravitational fields at the surfaces of these types of stars, for a typical range of masses. For the main sequence
stars we used the calculated Zero Age Main Sequence (ZAMS) masses and radii of the models in 
\cite{Schaller1992}, and for the white dwarfs we used the mass-radius
relation tabulated in \cite{Glendenning1996}.

\vspace{2mm}
\noindent {\em (iv)\/} The points labelled PSRs denote compact object binary systems in which at least one member is a pulsar. These points were calculated using the reduced mass and the semi-major axis of the binary as the relevant parameters in
equations (\ref{eq:potential}) and (\ref{eq:xi_astro}). Note that, interestingly, the curvature regime constrained by current measurements
of compact object binaries coincides with the regime constrained by near-Earth satellites. Therefore we should not be surprised to find that General Relativity, which describes near-Earth motions so flawlessly, also seems to be perfectly adequate for the inspiral phase of these binaries.

Compact object binaries can further test a phenomenon common to many alternative gravity theories: violations of the Strong Equivalence Principle (SEP). If a new field coupling to the matter energy-momentum tensor is present, its profile over the binary members can depend strongly on their internal structure \citep{TEGP, Will2006}. If the binary members have significantly different masses or compositions, they will experience different accelerations in an external gravitational field\footnote{This should also lead to the emission of dipolar gravitational radiation, which does not occur in GR.}. Attempts have been made to constrain this differential acceleration in the external gravitational field of the galaxy \citep{Stairs2005, Gonzalez2011}.

\vspace{2mm}
\noindent {\em (v)\/} The red points on the Milky Way (MW) contour mark
the gravitational fields probed by the S~stars orbiting close to the
black hole Sagittarius A$^{*}$. For these stars we used the orbital
properties given in Table~7 of \cite{Gillessen2009}, assuming a
central black hole mass of \mbox{$M=4.3\times 10^6 M_\odot$} and a distance
of $D=8.3$~kpc. 

\vspace{2mm}
\noindent {\em (vi)\/} The green points in the top right corner of Fig.~\ref{fig:parameter_space} represent a population of stellar mass black holes. We use the dynamical mass measurements reported in \cite{Ozel2010}, obtained from observations of transient low-mass X-ray binaries. $\xi$ and $\epsilon$ are evaluated at the radius of the innermost stable circular orbit (ISCO) for a Schwarzchild black hole\footnote{Recall that prefactors due to the rotation of the black hole are unimportant on the scale of our parameter space.}, as this is the smallest orbital radius from which we can obtain spectral information\footnote{Direct imaging of the photon ring would enable one to probe down to 1.5 Schwarzchild radii for a maximally rotating black hole, see \S\ref{sub:gaia}. }. 

\vspace{2mm}
\noindent {\em (vii)\/} The purple set of points on the far right of the plot represent supermassive black holes (SMBH), using mass measurements from \cite{Gultekin2009} and again evaluating the gravitational field at the ISCO.  Note that these all lie below the extended contour of the Milky Way (dashed), which intersects the purple point corresponding to Sagittarius A$^*$. This is a simple selection bias -- reliable mass measurements are easiest to make for the most massive extragalactic black holes, where the dynamical broadening of emission lines (used as a mass indicator) is most pronounced. 
\vspace{3mm}

\subsection{Galaxies and Clusters}
\label{sub:gal_cluster}

In contrast to the individual objects of \S\ref{sub:stellar}, galaxies and clusters probe a range of gravitational potentials and curvatures. The strong-field regions near the central black holes of galaxies can be treated as approximate vacuum populated by orbiting test bodies (see point \textit{(v)} above), and hence are represented as straight lines on our parameter space (see the discussion surrounding equation \ref{constM}).

However, the high-resolution observations needed to characterize the central region of a galaxy are currently only available for the Milky Way and the nearby massive galaxies such as M87. For the case of the Milky Way, we calculated the gravitational potential and
curvature near the central black hole using the mass profile inferred
by \cite{Schodel2002} from the orbits of nearby stars, and
smoothly connected it to the mass profile inferred by \cite{Dehnen1998} from the Galactic rotation curve. Similarly, for M87 we used the mass profile inferred by \cite{Gebhardt2009}
and smoothly connected it to the location of the event horizon of the
central black hole. 

Note that the physical size of the central galactic regions for the Milky Way and M87 are disproportionately over-represented in Fig.~\ref{fig:parameter_space}, because the curvature increases rapidly through many orders of magnitude there. Conversely, the (physically much larger) gradually-changing outer regions of the galaxies are represented by the non-straight sections of the MW and M87 lines at lower curvatures.

A further four lines in Fig.~\ref{fig:parameter_space} (dark blue) indicate the approximate gravitational fields that can be tested using galaxies other than the Milky Way and M87, for which only the bulk/outer regions are observable. We show data for two typical spiral galaxies (rightmost dark blue lines) and two dwarf galaxies (leftmost dark blue lines). We use rotation curves from The HI Nearby Galactic Survey (THINGS; \citealt{Blok2009}) and use the Keplerian scaling \mbox{$v^2\sim\Phi\sim GM/r$} to calculate the potential as a function of radius in these galaxies. The corresponding curvature at each radius is then simply related by a further factor of $\sqrt{48}/r^2$ (see equation \ref{eq:xi_astro}). 

We can see that one of the spiral galaxies is predominantly virialized, as it maps to a vertical line (see the discussion surrounding equation \ref{virial}). Interestingly, the two dwarf galaxies lie on and below the phenomenological acceleration-scale contour discussed in \S\ref{sub:understanding}, which roughly delineates where dark matter becomes a significant component of gravitational systems. It is well-known that the inferred mass-to-light ratios of dwarf galaxies are particularly large \citep{Walker2013}. This feature is borne out visually in the parameter space of Fig.~\ref{fig:parameter_space}. 

Three red lines in Fig.~\ref{fig:parameter_space} represent the galaxy clusters Abel~689, Abel~2219, and Abel~2261. Their mappings onto our parameter space were calculated in a manner entirely analogous to the galaxies, using the mass profiles inferred by \cite{Rines2013}. The curves are predominantly vertical, again reflecting the largely virial nature of clusters. 

We highlight that galaxy clusters have the unique property that they routinely cross the acceleration-scale contour. That is, their outskirts are strongly dark-matter dominated, but their cores are not. This makes them a potential testbed for non-standard dark matter theories; one might look for a evidence of a novel transition regime within them \citep{Lam2012}.

\subsection{Cosmological Quantities}
\label{sub:cosmo_lines}

Placing cosmological systems on Fig.~\ref{fig:parameter_space} requires a more subtle treatment than that of the previous subsections. We will make a split between the homogeneous cosmological background, described by the single function $\Hu(\eta)$, and the evolution of linear perturbations within the universe. \newline

\noindent \textbf{Background.}
Because the unperturbed FRW metric is isotropic, the unperturbed Kretschmann scalar does not contain any gradient terms; it is a function of time only. For convenience we repeat it here:
\begin{align}
\label{qem}
\0\xi^{\mathrm{cosmo}}(a)=\0K(a)&=\frac{\sqrt{12}}{a^2}\left(\dot\Hu^2+\Hu^4\right)^{\frac{1}{2}}\;.
\end{align}
We can use this quantity to assign a y-axis parameter to the homogeneous universe. Since a potential quantifies the difference between perturbations and a background, we cannot assign a $\epsilon$ quantity to an \textit{unperturbed} FRW metric. Hence we plot the curvature of the homogeneous universe as a horizontal line in Fig.~\ref{fig:parameter_space}.

The quantity $\0\xi^{\mathrm{cosmo}}$ moves vertically downwards as the universe expands and the conformal Hubble rate drops; the magenta lines in Fig.~\ref{fig:parameter_space} shows its value at a few key epochs. The uppermost magenta line shows $\0\xi^{\mathrm{cosmo}}$ at the time of Big Bang Nucleosynthesis (BBN), which -- like the spin-down rates of binary pulsar systems discussed previously -- coincides with the curvature scale probed by near-Earth and solar satellites. It is sometimes asserted that theories of modified gravity in the early universe are extremely tightly constrained by the success of our model of BBN (e.g.~\citealt{Kneller2003}). If corrections to the Einstein-Hilbert action are curvature-dependent then this argument does not apply: we should not be surprised that a description of gravity that works extremely well in the Solar System is also perfectly adequate for predicting the abundances of light elements \citep{Santiago1997}. 
 
In a similar vein, the lowest magenta line labelled `Lambda' represents the curvature of a FRW universe completely dominated by the cosmological constant. The value of $\0\xi^{\mathrm{cosmo}}$ for the \textit{present} universe sits a very small distance above the lambda line, because pressureless matter is not yet completely subdominant to the cosmological constant. The universe will asymptote towards the lambda line in the future, and this represents a fundamental minimum curvature scale. Clearly a universe with a different value of the cosmological constant would have a different minimum curvature.

The middle magenta line labelled `last scattering' represents the value of $\0\xi^{\mathrm{cosmo}}$ at $z\simeq1100$; the location of the CMB on the parameter space will be discussed shortly.
 
 Three important lines (nearly) intersect in the lower right-hand corner of the plot: the phenomenological acceleration scale, the lambda line, and the horizon boundary. This is an incidental manifestation of the coincidence problem, as follows: recall from \S\ref{sub:understanding} that the phenomenological value of the acceleration scale below which we need to invoke the existence of dark matter is roughly coincident with the value of $c\times H_0$. This acceleration scale then determines $\Omega_{CDM}$, which together with $H_0$ determines $\Omega_\Lambda$  (via the Friedmann equation). Because $\Omega_{CDM}$ and $\Omega_{\Lambda}$ are of comparable size today, the lambda line and acceleration line intersect. Since the naive `Schwarchild radius of the universe'\footnote{One can perform an extremely naive calculation that treats the universe as a sphere with uniform density \mbox{$3 M_P^2 H_0^2$}, and radius \mbox{$R_U\sim c/H_0$} (based only on dimensional arguments and characteristic quantities). This leads to the conclusion that the Schwarzchild radius of the universe -- if one could define such a thing -- is its own horizon scale, $c/H_0$. Hence horizon scales are inherently relativistic.} is roughly $c/H_0$, it is no surprise to find that these three components converge in the lower right of the plot.
 
\vspace{3mm} 
  
\noindent\textbf{Perturbations.} Most observables of interest in cosmology depend on perturbations of the spacetime, for example, galaxy clustering, galaxy weak lensing, the CMB and CMB lensing. Information about the amplitudes and scales of density perturbations in the universe is encapsulated in the matter power spectrum \citep{Tojeiro2014}. In \S\ref{sub:cosmo_params} and Appendix \ref{app:gi_cosmo} we explain how to quantify the potentials and curvatures probed by these density perturbations in terms of our parameters $\epsilon^{\mathrm{cosmo}}(k,a)$ and $\xi^{\mathrm{cosmo}}(k,a)$. 

We will use only the linear matter power spectrum, which clearly is not accurate at small scales\footnote{Of course it is the complicated evolution of small-scale density perturbations which leads to the richness of the overall parameter space.}. The cyan curve in Fig.~\ref{fig:parameter_space} (lower section, solid) represents the \textit{perturbative} piece of the Kretschmann curvature, shown here at redshift $z=0$. Note that to obtain the y-axis values of this curve we have subtracted the zeroth-order contribution due to the homogeneous expanding universe, $\0\xi^{\mathrm{cosmo}}(a)$, because otherwise the perturbations would be imperceptible on the diagram. Recall our statement from \S\ref{section:quant} that this is conceptually analogous to the study of CMB temperature fluctuations: the temperature monopole of $\sim 2.725$K is subtracted, and the perturbations about this mean temperature are studied.

As density perturbations collapse they shrink in physical size and deepen in amplitude, increasing the $\epsilon^{\mathrm{cosmo}}$ and $\xi^{\mathrm{cosmo}}$ values that they probe. This means that cosmological perturbations which have decoupled from the background expansion move \textit{upwards} in Fig.~\ref{fig:parameter_space}. Perturbations which have not yet decoupled will be dragged overall \textit{downwards} by the decreasing value of  $\0\xi^{\mathrm{cosmo}}(a)$ (equation \ref{qem}).

The `knee' in the cyan curve corresponds to the turnover in the matter power spectrum, with large distance scales to the \textit{right} of the knee (note that $P_M(k)$ gets `flipped' horizontally with respect to standard plots of the power spectrum, when mapped onto our parameter space). The Baryon Acoustic Oscillations (BAO) can just about be discerned to the \textit{left} of the knee. The largest scales asymptote to a potential value of $\sim 10^{-5}$, in agreement with the amplitude of fluctuations seen in the CMB. Having entered the cosmological horizon only recently, these perturbations have had little time to evolve from their `frozen' super-horizon values \citep{peacock1990}.
\vspace{3mm} 

\noindent\textbf{Cosmic Microwave Background.} The angular scales and amplitudes of the first few CMB peaks have been measured to high precision by the experiments WMAP and Planck \citep{Bennett2013,Planck2014}. To determine the implications of these measurements for our parameter space, we need to assign scale-dependent $\epsilon$ and $\xi$ parameters to the spectrum of potential wells present \textit{during the recombination era}. Although we introduced similar quantities earlier in this subsection and in Appendix~\ref{app:gi_cosmo}, the derivation there made use of the matter power spectrum to provide information about clustering on different scales. As such, those quantifiers were appropriate for tests of gravity using large-scale structure \textit{at late times} (e.g. growth statistics and weak lensing).

At the time of recombination, when little structure formation has occurred, it is more appropriate to glean scale-dependent information from the primordial power spectrum laid down during inflation. It is also more helpful to work in terms of angular modes $\ell$ than in wavenumbers $k$; we can then straightforwardly extract information relevant to the CMB peaks by evaluating our expression at $\ell\simeq 222,\,537,\,816\ldots$, i.e. the $\ell$-values of the first few CMB peaks. Our derivation assumes that the primordial power spectrum has a power-law form, but is otherwise independent of the specific details of the inflationary mechanism.

The derivation of $\epsilon$ and $\xi$ for the CMB peaks is presented in Appendix~\ref{app:CMB}. Here we will show only the final results ($\Xi$ was defined in equation \ref{kgi}):
\begin{align}
\epsilon_\ell&=\sqrt{\frac{2}{\pi}\int dk\,k^2\,|\Phi_\ell(k)|^2\,P_{0}(k)}\\
 \xi_\ell &= \sqrt{\frac{2}{\pi} \int dk\, k^2 |\Xi_\ell(k)|^2 \, P_{ 0}(k)}\;,
\end{align}
where
\begin{align}
\Phi_\ell(k)&=\int_0^{\eta_0} d\eta\, j_\ell\left[k(\eta_0-\eta)\right]  \,g_{\mathrm{vis}}(\eta) \,\Phi(k,\eta) \label{epsl}\\
\Xi_\ell(k)&=\int_0^{\eta_0} d\eta\, j_\ell\left[k(\eta_0-\eta)\right]   \,g_{\mathrm{vis}}(\eta) \,\Xi(k,\eta)\;.\label{xil}
\end{align}
$P_0(k)$ is the primordial power spectrum and $j_\ell(x)$ are the spherical Bessel functions, and the normalized visibility function $g_{\mathrm{vis}}(\eta)$ is peaked at the epoch of recombination. Through this choice of window function we are focussing on primary anisotropies. Secondary anisotropies induced by, e.g. the Intergrated Sachs-Wolfe effect and CMB lensing, do not significantly impact Fig.~\ref{fig:parameter_space}.

The green markers in Fig.~\ref{fig:parameter_space} (lower middle) show the results of the calculations for the first four CMB peaks, evaluated using Planck 2013 cosmological parameters. Their location agrees with expectations based on extrapolating the cyan curve of the matter power spectrum (see earlier in this subsection) to higher redshifts. 

However, the coincidence of the CMB points with the gravitational fields of galaxies and clusters is (perhaps) unexpected. It suggests that if we wish to stringently test for deviations from $\Lambda$CDM+GR, simply picking seemingly very different phenomena -- such as galactic rotation curves and the CMB -- is insufficient. We need to think more carefully to ensure that we are not in fact always testing gravity in the same few regimes.

The observant reader may note that the CMB peak points sit above the phenomenological acceleration scale. In \S\ref{sub:understanding} we stated that it was \textit{below} this line that dark matter becomes highly relevant; yet dark matter is essential for correct modelling of the CMB temperature power spectrum. The simple explanation for this is that the phenomenological acceleration scale shown  in Fig.~\ref{fig:parameter_space} uses \textit{today's} Hubble constant, i.e. it is $c \,H_0$. If we plotted the acceleration scale $c\,H(z=1100)$, the CMB peaks would lie below it. In any case, until we have a better understanding of the nature of dark matter, the extrapolation of this phenomenological acceleration scale to earlier epochs is a largely speculative exercise.
\vspace{3mm}

Before concluding this subsection, it is worth highlighting that the CMB peaks and the horizontal BBN and last scattering lines are probes of gravitational fields at \textit{early} cosmological times. Many alternative theories of gravity are deliberately constructed to deviate from GR at late times in the universe. So, although GR has effectively been tested at fields of $\{\epsilon,\,\xi\}\simeq\{10^{-12},\,10^{-29}\}$ (by the successes of BBN models) at early epochs, we cannot claim to have tested this location of the parameter space in the late-time universe.
 
 \subsection{A Note of Caution}
\label{sub:caution}
Placing all gravitational systems in the universe on a single diagram is made difficult by the dynamic range involved. The  satellite experiments, planets, stars and binary pulsars we use to test gravity all sit inside the Milky Way\footnote{As do the compact objects, but their values of $\epsilon$ dominate the Galactic contribution locally anyway.}, and make only tiny perturbations to the potential profile of the galaxy. One could argue, then, that all of these systems should simply be placed at a potential of $\epsilon\sim 10^{-6}$ (the approximate value of $\epsilon$ in the outer parts of the Galaxy). However, this would lead to a very unilluminating plot.

 We have avoided this problem by considering each astrophysical system in isolation. That is, we consider the planets moving in the potential of the Sun and take the metric to be Minkowskian at larger distances, ignoring the presence of the Galaxy. This is acceptable because the Galactic potential varies negligibly over the scale of the Solar System, so it becomes a constant, unmeasurable `zero-point' contribution. Similarly, for near-Earth satellite experiments we consider the satellites to be test particles orbiting the Earth and ignore the gravitational field of the Sun. 

One could draw an analogy with the manner in which a cosmological overdensity with wavelength greater than the horizon ceases to be regarded as a perturbation, and instead simply shifts the mean matter density of the observable universe.

It is for this reason that the satellite experiment point ($\xi\sim 10^{-27}$) does not sit on the Solar System (SS) line of points (dark blue circles). Similarly, the S stars have a dual presence on the Main Sequence curve (MS, red curve).

One disadvantage of this approach is that it might render our gravitational parameter space unsuitable for analyzing some kinds of screening mechanisms in alternative theories of gravity. For example, in theories that exhibit chameleon screening, one expects systems with a gravitational potential $\Phi\gtrsim 10^{-6}$ to be screened \citep{Jain2011, Jain2013} i.e., deviations from GR become suppressed there. The chameleon mechanism is thought to be sensitive to the \textit{total} gravitational potential present, including background contributions. If this is true, then the planets of our Solar System should be screened by the surrounding potential of the Galaxy, even though on our plot they sit at potentials smaller than $\epsilon=10^{-6}$.

The goal of this paper is to unify tests of GR, and Fig.~\ref{fig:parameter_space} is consistent with these aims. It is quite possible that a different choice of axes (perhaps $\epsilon$ vs. a measure of mean density) would lead to a parameter space more suitable for screening applications.

\section{Experimental Tests of Gravity}
\label{section:experiments}
Thus far we have developed a framework for teasing apart different gravitational fields that exist in the universe. The next natural questions to ask are: which of these gravitational fields have been probed experimentally to date? Which might we hope to measure in the near future? For observers, such regions are of interest to ensure that constraints on deviations from $\Lambda$CDM+GR are watertight, with all bases checked. For model-building theorists, these remaining regions should provide guidance for the kinds of gravity theories that are still viable.

Fig.~\ref{fig:exp} indicates the regions of our gravitational parameter space probed by different experiments; note that the majority of the curves shown represent future experiments. Notice also that different experiments probe different aspects of gravitational fields, e.g. the gravitational wave experiments probe the dynamics of spin-2 perturbations, the CMB and galaxy surveys probe the dynamics of spin-0 perturbations, and laboratory tests generally probe static gravitational fields. It is possible for deviations from GR to be manifest in one of these situations but not in others. For example, the Kerr solution is common to many theories of gravity \citep{PsaltisPerrodin2008}, but the dynamics of spin-2 perturbations about the Kerr solution are not \citep{Barausse2008}.

Below we will explain the specifics of the boundaries/points plotted, but let us first summarize the significant implications of this diagram.

The most striking feature of Fig.~\ref{fig:exp} is that there is a `desert' between curvatures of $\sim 10^{-38}-10^{-50}$ cm$^{-2}$, where there are no tests of GR. Comparing to Fig.~\ref{fig:parameter_space}, we see that there is a distinct shortage of gravitational systems with which to test these fields. The only systems spanning the untested window, in theory, are galaxies: the desert straddles the region where their rotation curves transition from the inner Schwarzchild-like orbits dominated by the central black hole to the outer regions dominated by dark matter. Unfortunately, for anything except the Milky Way and M87, we can only observe the outer regions due to resolution limits\footnote{Recall that the extent of the lines representing the inner/outer regions on  Fig.~\ref{fig:parameter_space} does not reflect their relative physical sizes.}.

Furthermore, the untested desert sits between a region where GR is extremely well-constrained (the Solar System, binary pulsars, etc.) to the `problem area' at the bottom of the plot, where dark matter and dark energy are invoked to fit observations. An alternative to invoking dark energy is that within this untested desert there exists a transition scale marks the onset of corrections to GR \citep{BakerQS}.

The PPN formalism (see \S\ref{sub:PPN} for details) tests the form of the metric when expanded in powers of \mbox{$\Phi\sim v^2/c^2$}. As such, it is designed to test along the x-axis direction of our parameter space, but has relatively little power to probe the y-axis direction. Needless to say, this is why cosmological modified gravity is not ruled out by Solar System constraints, even in the absence of an explicit screening mechanism. At least in the quantification scheme developed here, the relevant gravitational fields are separated by many, many orders of magnitude in curvature. This leaves ample room for new physics to come into play.

\begin{center}
\begin{figure*}[t]
\begin{center}
\psfig{file=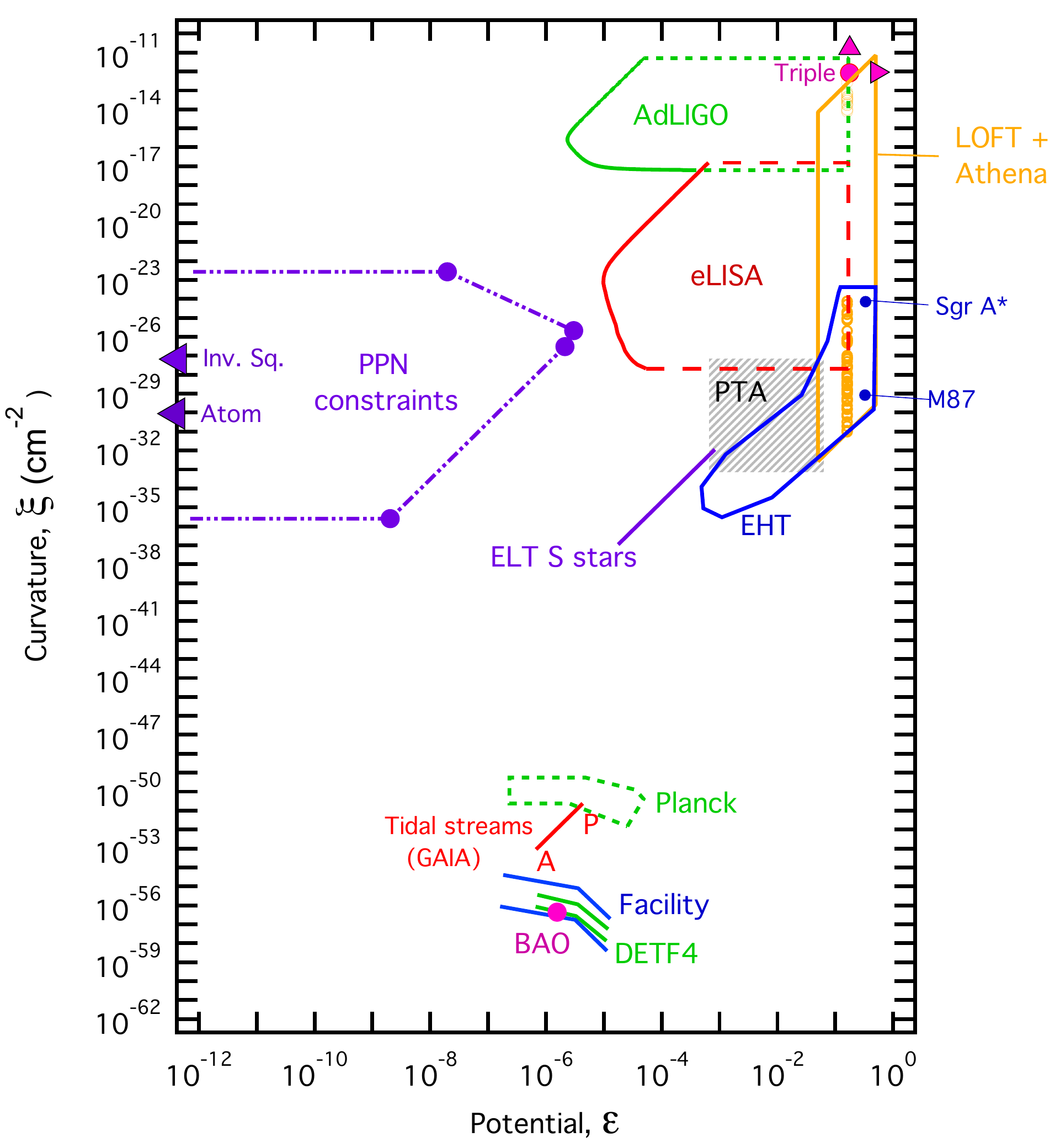,width=15cm}
\caption{The experimental version of the gravitational parameter space (axes the same as in Fig.~\ref{fig:parameter_space}). Curves are described in detail in the text (\S\ref{section:experiments}). Some of the abbreviations in the figure are: PPN = Parameterized Post-Newtonian region, Inv. Sq. = laboratory tests of the $1/r^2$ behaviour of the gravitational force law, Atom = atom interferometry experiments to probe screening mechanisms, EHT = the Event Horizon Telescope, ELT = the Extremely Large Telescope, DETF4 = a hypothetical `stage 4' experiment according to the classification scheme of the Dark Energy Task Force \citep{Albrecht2006}, Facility = a futuristic large radio telescope such as the Square Kilometre Array.}
\label{fig:exp}
\end{center}
\end{figure*}
\end{center}

\pagebreak
\subsection{Cosmology}
\noindent{\bf Galaxy Surveys.} In the lower section of the figure we indicate the regions probed by two future galaxy clustering surveys. In green we consider a next-generation `stage 4' space-based survey of the kind envisaged by the Dark Energy Task Force \citep{Albrecht2006}, labelled DETF4. In blue, we consider a futuristic `Facility stage' ground-based radio interferometer of the kind considered by \cite{Bull2014}, capable of mapping nearly the full sky out to very high redshifts.

Each survey is delineated by two lines, whose separation is set by the survey redshift range. We used equations (\ref{epsilon_cosmo}) and (\ref{xi_cosmo}) to plot the minimum and maximum $k$-values for each experiment, where the minimum $k$ is set by the size of the survey and the maximum $k$ is chosen to cut off before nonlinearities become dominant (the value chosen varies somewhat in the literature for the different experiments). We have also plotted a point of \mbox{$k\simeq 0.05$ h Mpc$^{-1}$}, corresponding to the approximate position of the turnover in the matter power spectrum. The bent shape of these survey regions reflects the shape of the matter power spectrum shown in Fig.~\ref{fig:parameter_space} (cyan curve). Table~\ref{tab:surveys} shows the values used. In addition, we have added a point to represent recent measurements of the BAO feature \citep{SDSSBAO}.

Although the extent of the parameter space probed by cosmology is small, we stress that this is one of the most crucial regions of the plot. Indeed, it is only via cosmology that we are able to access the ultra-low curvatures where the problematic dark sector(s) are completely dominant.
\vspace{2mm}

\begin{deluxetable}{ccccc}
  \tablecolumns{5}
  \tablewidth{0pt}
\tablecaption{Galaxy Survey Parameters}
  \tablehead{\colhead{Experiment} & \colhead{$k_{low}$ (h Mpc$^{-1}$)}  & 
\colhead{$k_{high}$ (h Mpc$^{-1}$)}  & \colhead{$z_{low}$} & 
\colhead{$z_{high}$ }}
\startdata
DETF4 & 0.006 & 0.2 & 0.65 & 2.05 \\
Facility & 0.004 & 0.5 & 0.42 & 7.0 \\
BAO & - & 0.1 & - & 0.57 
\enddata
\label{tab:surveys}
\end{deluxetable}

\noindent{\bf The Cosmic Microwave Background.} The green dashed region in Fig.~\ref{fig:exp} denotes the constraints from the ESA Planck satellite \citep{Planck2014}. This corresponds to the location of the peaks in the CMB temperature power spectrum, as shown in Fig.~\ref{fig:parameter_space}. The contour has been extended to slightly smaller potentials than indicated in Fig.~\ref{fig:parameter_space} to incorporate $\ell$-values beyond the first four peaks (note that, analogous to the matter power spectrum, the largest scales of the CMB are at the right-hand end of the Planck region).

We remind the reader of the comment made at the end of \S\ref{sub:cosmo_lines}: that the primordial CMB anisotropies probe gravity at a different cosmological time to most of the other curves on the parameter space. To test time-dependent modifications to gravity\footnote{To give a very simple example (not necessarily the best-motivated one), think of `freezing' or `thawing' quintessence models (see \citealt{Tsujikawa2013} and references therein). In these theories the time-dependence of deviations from GR comes from a specific choice of potential for the scalar field. In this case the deviations are not \textit{directly} dependent on the ambient Kretschmann curvature.}, one might wish to add a third axis to the plot, indicating the redshift at which a test of gravity takes place. The constraints shown in Fig.~\ref{fig:exp} would then occupy volumes in this three-dimensional parameter space. A theory-specific deviation from GR would also carve out a volume of this space, as a function of redshift. The question of interest would be the extent of intersection between this volume and those of the gravity tests.

\subsection{Galactic-Scale Tests of Gravity}
\label{sub:gaia}

\noindent {\bf Tidal Streams.}
With the 2013 launch of the ESA Gaia mission \citep{GAIA2001}, we will soon be in possession of much-improved data on the orbits and extents of several tidal streams of the Milky Way, which act as probes of the Galactic density profile. The sophisticated orbit modelling applied to Gaia data (e.g. \citealt{Binney2005}) requires a gravity theory as an input to calculate precise relativistic effects.

We are not currently aware of any initiatives to test GR with Gaia's tidal stream data; however, they provide an opportunity to probe a severely under-populated region of the gravity parameter space at late cosmological times (see \citealt{Penarrubia2012} for related ideas). On Fig.~\ref{fig:exp} we have drawn a line to indicate the range of potentials and curvatures that could be probed using tidal streams, using the smallest pericentre distance of 14 kpc (belonging to the GD-1 stream) and the largest apocentre of 90 kpc (belonging to the Orphan stream, see \citealt{Newberg2009,Law2010, Koposov2010}). We have not plotted separate lines for each tidal stream because they would be indistinguishable on the scale of the diagram. The labels `A' and `P' indicate the apocentre and pericentre respectively.

Our estimates here are approximate because we have not accounted for the non-sphericity of the Milky Way. Neither have we used an accurate density profile to calculate the fraction of the Milky Way's mass interior to the orbits. The effects of both these corrections is likely to be negligible given the logarithmic axes of the figure.\newline

\noindent{\bf The Event Horizon Telescope.} 
The Event Horizon Telescope (EHT) is a network of millimetre and sub-millimetre telescopes being used for Very Long Baseline Interferometery (VLBI) to directly image supermassive black holes at the centre of galaxies. The prime targets of the EHT are Sagitarius A$^*$ and M87, for which early observations have shown that
the size of their emitting regions at a wavelength of 1.3 mm is comparable to their corresponding Schwarzschild radii
\citep{Doeleman2008,Doeleman2012}. Future observations will reveal the `shadows' of the black holes
against a bright photon ring (that lies at 1.5 Schwarzchild radii for a non-rotating black hole). Naturally the EHT can study other central black holes as well, but not at sufficient resolution to directly image the event horizon.

Two blue markers to the right-hand side of Fig.~\ref{fig:exp} show the gravitational fields probed by these observations of Sgr A$^*$ and M87. They lie to the right of the yellow LOFT+Athena points (to be discussed shortly), which were evaluated at the ISCO (3~R$_S$). Neglecting the effects of rotation is an acceptable simplification for the purposes of this paper. Note that the ELT S stars line (purple), when extended, would reach the Sgr A$^*$ point, as expected for a constant mass contour (see \S\ref{sub:understanding}).
 
The blue bounded region on the mid-right of Fig.~\ref{fig:exp} shows the gravitational fields probed by other EHT sources. This region was determined by using mass and redshift estimates for the sources, and calculating the expected resolution of the EHT at that redshift in units of the Schwarzchild radius of the black hole. For example, the resolution of the EHT at 4.4 Mpc is $\sim 15$ times larger than the event horizon of Centaurus A, based on a mass estimate of $\sim 3\times10^8$ M$_\Sun$ \citep{Johannsen2012SMBH}. Hence for Centaurus A we evaluate $\epsilon$ and $\xi$ (equations~\ref{eq:potential} and \ref{eq:xi_astro}) at 15~R$_S$ from the black hole. This is why the EHT region extends to the left of the LOFT+Athena region: the fields probed by the EHT are at greater distances from the central source.

\subsection{S Stars}
Fig.~\ref{fig:exp} bears a line marked `ELT S stars', which represent improved observations of stars in close orbits around Sgr A$^*$ with a next-generation large telescope (anticipated aperture $\sim$ 30m). The calculation of these points is analogous to the S stars discussed in \S\ref{sub:stellar}. The range of pericentres was taken from the simulations of \cite{WeinbergELT}.

\subsection{The PPN Regime}
\label{sub:PPN}
Arguably the most stringent tests of GR to date are those made using the Parameterized Post-Newtonian formalism (PPN), a long-established framework for testing weak-field gravity on small scales\footnote{The coordinate system used in PPN requires that the metric is Minkowskian at large distances from the system in question, see p92 of \cite{Will2006}. This choice implies that the system is small enough that cosmological expansion can be ignored. (Of course, it is recognized that the PPN coordinates must ultimately match onto the cosmological solution at even larger distances).} (\citealt{Will2006} and references therein). PPN proceeds as an expansion of the metric about a Newtonian background in terms of small quantities of order \mbox{$\Phi\sim v^2/c^2$}, where the expansion terms describe corrections to Newtonian gravity. The formalism delivers a set of ten convenient parameters  quantifying different physical effects that could potentially indicate deviations from GR.

Of these ten parameters, one of the most commonly discussed is denoted by $\gamma$. Loosely speaking, $\gamma$ quantifies how responsive the curvature of spacetime is to mass. The tightest constraint on $\gamma$ to date comes from measurements of the Shapiro time delay effect by the Cassini spacecraft, and stands at \mbox{$|\gamma-1|=(2.1\,\pm\,2.3)\times 10^{-5}$} \citep{Bertotti2008}. The parameter $\gamma$ has also been constrained from the gravitational lensing of radio sources by the Sun \citep{Shapiro2004}. 

The next foreseeable improvement in constraints on $\gamma$ is expected to come from the ESA Gaia experiment introduced in \S\ref{sub:gaia}. Using Gaia's highly precise astrometric data to measure light deflection by both the Sun and Jupiter, constraints as tight as \mbox{$|\gamma-1|\lesssim 5\times 10^{-7}$} might be possible \citep{GAIA2001}\footnote{There is some uncertainty over the precise bound at present, due to a stray light problem found during the commissioning phase of Gaia.}.

Drawing together the results above, we have marked on Fig.~\ref{fig:exp} the approximate region of our parameter space to which the PPN formalism has been applied. The four purple points (filled circles) mark, clockwise from the uppermost, 
the Gaia lensing constraint, the Double Pulsar, the Sun and tests performed using satellites in the Solar System. Note that Jupiter effectively appears twice: the uppermost purple point marks the gravitational fields close to the surface of Jupiter, as probed by the Gaia lensing measurement. The lowermost purple point marks the conditions near the mid- and outer planets, when considered as test particles moving in the gravitational field of the Sun. This regime is probed by the transmission of signals between satellites in the outer parts of Solar System and Earth\footnote{The classic example is measurements of the Shapiro time delay \citep{Shapiro1964,Bertotti2008}, though in that case most of the delay is accrued when the photons are passing close to the Sun.}.

Though not strictly an application of PPN, we have used laboratory experiments (\S\ref{sub:laboratory}) as a sensible boundary for the minimum potentials that have been stringently tested -- hence the extension of the PPN region to the x-axis.

\subsection{Laboratory Tests of Gravity}
\label{sub:laboratory}
A number of gravity theories predict the gravitational force law to deviate from the Newtonian $1/r^2$ dependence at short scales (see \citealt{Adelberger} and references therein). Laboratory experiments have been performed that aim to constrain deviations of the force law between two masses $m_1$ and $m_2$, parameterized in terms of a Yukawa potential,
\begin{align}
\label{Yukawa}
V(r)&=-\frac{Gm_1 m_2}{r}\left(1+\alpha e^{-\frac{r}{\lambda}}\right)\,.
\end{align}
Constraints are placed on the strength parameter $\alpha$ at a given experimental length scale $\lambda$ (see Fig.~4 of \citealt{Yang2012}).

\cite{Kapner2007} found that any new gravitational-strength force ($|\alpha|\simeq 1$) must have an interaction length $\lambda \leq 56\mu m$). This constraint was obtained using an E\"{o}t-Wash experiment to measure the gravitational
acceleration between two large disks of mass density \mbox{$\rho= 10.3$~kg~m$^{-3}$} and thickness $h= 1$~mm. In Newtonian theory the gravitational acceleration between the two disks becomes independent of their separation at short distances, and is given approximately by
\begin{align}
a_{\mathrm{Eot}}=2\pi G \rho h \simeq 4.3\times 10^{-10}~\mbox{cm}~\mbox{s}^{-2}\,.
\end{align}
Hence the corresponding potential and curvature probed by this experiment are:
\begin{align}
\epsilon_{\mathrm{Eot}}&\simeq \left(\frac{a_{\mathrm{Eot}} \lambda}{c^2}\right) \simeq 2.7\times 10^{-33}\\
\xi_{\mathrm{Eot}}&\simeq\frac{\sqrt{48}\,\epsilon_{\mathrm{Eot}}}{\lambda^2}\simeq 5.9\times 10^{-28}~\mbox{cm}^{-2}\,.
\end{align}
Due to the very small value of $\epsilon_{\mathrm{Eot}}$, we indicate this experiment by a marker labelled `Inv.~Sq' on the y-axis of Fig.~\ref{fig:exp}.

\cite{Burrage2014} have recently proposed a different laboratory test of gravity, specifically designed to place constraints on the chameleon screening mechanism. They propose to use atom interferometry to measure the difference in gravitational potential felt by atoms that travel different paths in a gravitational field. In a vacuum chamber of the correct size these atoms would be unscreened, so any gravitational fifth forces would contribute to the potential they experience.

The region of our parameter space that this experiment probes can be simply estimated using equations (\ref{eq:potential}) and (\ref{eq:xi_astro}). For a chamber of $\sim 10$cm diameter and a central source mass of density $\rho\sim 1$g~cm$^{-3}$ and radius $1$cm, the relevant values are \mbox{$\epsilon_{\mathrm{atom}}\simeq3.1\times10^{-29}$} and \mbox{$\xi_{\mathrm{atom}}\simeq2.2\times10^{-30}$}. Similar to the inverse square law tests, this experiment is marked as a point on the y-axis of Fig.~\ref{fig:exp}.

Further recent work related to laboratory tests of chameleon screening can be found in \cite{BraxDavis2014}.

\subsection{Gravitational Waves}
\label{sub:grav_waves}
\noindent{\bf Interferometers.} The planned AdLIGO and eLISA experiments have the potential to make the first direct detections of gravitational waves \citep{LIGO2013,eLISA2013}. The AdLIGO detector is due to come online in 2015, and is frequency-optimized to detect gravitational waves emitted during the final stages of the inspiral and coalescence of compact object binaries. The space-based eLISA mission is scheduled for launch in $2034$, and is sensitive to gravitational waves from the coalescence of supermassive black holes.

To determine the precise region of our parameter space that AdLIGO and eLISA probe, we make use of the \textit{inspiral horizon distance}, defined by \cite{LIGO2012}. Normally used as an indicator of instrument performance, the inspiral horizon distance measures the maximum distance at which the coalescence of an equal-mass compact object binary could be cleanly detected (signal-to-noise ratio, SNR $\gtrsim 8$). The quantitative definition of the inspiral horizon distance folds in information about both the expected waveform of the coalescence and the frequency-dependent sensitivity of the interferometer:
\begin{align}
\label{wkt}
D_{IHD} &= \frac{1}{8}\sqrt{\frac{5\pi }{24c^3}}(G\mathcal{M})^{5/6}\pi^{-7/6} \sqrt{4 \int_{f_{low}}^{f_{high}} \frac{f^{-7/3}}{S_n(f)}df }\;,
\end{align}
where $D_{IHD}$ is the inspiral horizon distance, $S_n(f)$ is the power spectral density of the noise in the detector (in units of Hz$^{-1}$) and $\mathcal M$ is the chirp mass of the binary, related to the reduced mass $\mu$ and the total mass $M_T$ by
 \begin{align}
 \label{iaj}
 {\mathcal{M} }&=\mu^{\frac{3}{5}}M_T^{\frac{2}{5}} \;.
 \end{align}
 The frequency $f_{low}$ in equation (\ref{wkt}) is the low-frequency limit at which the detector noise becomes problematic ($\sim 10\,$Hz for AdLIGO and $\sim 10^{-5}\,$Hz for eLISA); $f_{high}$ will be defined momentarily.
 
 We will manipulate this definition of $D_{IHD} $ for our purposes. Rather than a distance measure, what we are seeking is a relationship between potential and Kretschmann curvature for systems detectable by AdLIGO and eLISA. Therefore we will set $D_{IHD}$ to be a typical source distance for these experiments (we use 12\,Mpc for AdLIGO and 2\,Gpc for eLISA) and use equation (\ref{wkt}) to find a relation between $\epsilon$ and $\xi$ for gravitational waves. Note that we are applying the inspiral horizon distance statistic to eLISA, though it was originally introduced for ground-based experiments. The implicit assumption here is that, to leading order, the \textit{shape} of the signal produced by the coalescence of supermassive black holes (eLISA sources) is similar to that of stellar mass coalescences (AdLIGO sources), just shifted to lower frequencies.
 
 First we rearrange equation (\ref{wkt}) to:
 \begin{align}
 GM &= \left(\frac{D_{IHD}}{B}\right)^{\frac{6}{5}} 2^{\frac{1}{5}} \left[4 \int_{f_{low}}^{f_{high}} \frac{f^{-7/3}}{S_n(f)}df \right]^{-\frac{3}{5}}
 \label{qef}
\end{align}
where we have considered an equal-mass binary system, made use of equation (\ref{iaj}), and bundled up some of the constants as
\begin{align}
\label{Bdef}
B=\frac{\pi^{-\frac{7}{6}}}{8}\left(\frac{5\pi}{24 c^3}\right)^{\frac{1}{2}}\;.
\end{align}
Close to one of the binary companions, we can take the Schwarzchild expressions for $\epsilon$ and $\xi$ to hold at leading order. Formally, we should use the formalism of \cite{Buonanno1999} to map the binary dynamics onto an effective one-body problem, and use the appropriate reduced mass quantities in our expressions. However, maintaining our policy of neglecting coefficients of order unity, we will ignore these factors here. 

We combine equations (\ref{eq:potential}) and (\ref{eq:xi_astro}) to obtain an expression for the orbital radius:
\begin{align}
\label{iop}
r&=(48)^{\frac{1}{4}}\sqrt{\frac{\epsilon}{\xi}}\;.
\end{align}
Then dividing equation (\ref{qef}) by equation (\ref{iop}), we obtain
\begin{align}
 \frac{GM}{r c^2} &=\epsilon= \left(\frac{D}{B}\right)^{\frac{6}{5}} \frac{2^{\frac{1}{5}}}{c^2} \left[4 \int_{f_{low}}^{f_{high}} \frac{f^{-7/3}}{S_n(f)}df \right]^{-\frac{3}{5}}\frac{1}{(48)^{\frac{1}{4}}}\sqrt{\frac{\xi}{\epsilon}}\;.
\end{align}
Finally, rearranging for $\epsilon$:
\begin{align}
\epsilon&= \left(\frac{D}{B}\right)^{\frac{4}{5}} \left(\frac{2^{\frac{1}{5}}}{c^2}\right)^{\frac{2}{3}} \left[4 \int_{f_{low}}^{f_{high}} \frac{f^{-7/3}}{S_n(f)}df \right]^{-\frac{2}{5}}\frac{1}{(48)^{\frac{1}{6}}}\xi^{\frac{1}{3}}\;.
\label{jos}
\end{align}
All that remains is to choose the upper limit of the frequency integral. During the inspiral phase of a binary, the instantaneous gravitational wave frequency is twice the instantaneous orbital frequency, which means that it can be (approximately) related to the Kretchmann curvature as
\begin{align}
f&=\frac{2}{P_{\mathrm{orb}}}=\frac{\sqrt{GM}}{\pi r^{\frac{3}{2}}}=\frac{c}{48^{1/4}\pi}\xi^{1/2}\;.
\label{tel}
\end{align}
We use the above expression to convert the frequency range of each experiment into a range of $\xi$. We then consider each $\xi$ in this range in turn, plugging it into equation (\ref{jos}) and setting the upper limit in the integral to be the corresponding $f$ given by equation (\ref{tel}); this yields a range of $\epsilon$ values. Effectively, we are calculating the total integrated signal an inspiralling system will have accumulated in the detector during its evolution up to that frequency (since frequency increases as the inspiral progresses).

The results of our calculations are shown in the solid bands labelled `AdLIGO' and `eLISA' in Fig.~\ref{fig:exp}. Events detected with SNR $>8$ occupy the area to the right of the solid bands; hence these parts of the parameter space can be probed. The upper limit of the potentials probed corresponds to conditions reached at the ISCO. The Kretschmann curvatures that can be probed is set by the frequency range of the experiments via equation (\ref{tel}). 

The shape of the detector sensitivity function of AdLIGO and eLISA is the reason for the bow-like shape of the solid curves. Once the value of $f_{high}$ exceeds the frequency of peak detector sensitivity the integral in equation (\ref{jos}) stays roughly constant, so that we have a curve with the approximate form $\xi\propto \Phi^3$. This is why the solid bands tend to straight lines in their upper regions.

We note that the eLISA region only encapsulates about half of the SMBH points. This is due to the design changes from the original LISA mission concept (which would have covered nearly all the SMBH points) to eLISA, which has shorter interferometer arms and hence is sensitive to higher frequencies. However, from the perspective of Fig.~\ref{fig:exp} this design change is unproblematic: the eLISA region still bridges the PPN region and the strong-field AdLIGO region, giving a more complete coverage of the parameter space.
 \vspace{2mm}

\noindent{\bf Pulsar Timing Arrays.} 
The passage of a gravitational wave between Earth and a pulsar will induce a shift in the pulse arrival time. By measuring   correlated shifts within a network of pulsars, pulsar timing arrays (PTAs) aim to detect gravitational waves via this effect.  Millisecond pulsars are used because they suffer smaller intrinsic timing irregularities.

Three such pulsar monitoring networks are currently operation: the Parkes Pulsar Timing Array \citep{Hobbs2009}, the European Pulsar Timing Array \citep{EPTA2010}, the North American Nanohertz Observatory for Gravitational Waves \citep{Jenet2009}, with future plans to combine all three networks to form the International Pulsar Timing Array \citep{Hobbs2010}.   
 
Due to the different detection method, it is not possible to apply the expressions we used for AdLIGO and eLISA to PTAs (note that equation \ref{wkt} required knowledge of the interferometer noise curve $S_n(f)$, which is not appropriate for a PTA). Development of rigorous $\epsilon$ and $\xi$ statistics for PTAs is beyond the scope of this paper; instead, we will settle for approximate numbers, as follows. We take the characteristic frequency range of PTAs to be $10^{-9}-10^{-6}~$Hz, and use equation (\ref{tel}) to convert this into a range of curvature. We then use equation (\ref{constM}) to calculate the corresponding range of potentials for a constant central mass $M$. We take the value of $M$ to be $10^9~$M$_\Sun$, representative of a large SMBH. Note that, as justified below equation (\ref{Bdef}), we are using the expressions for an isolated body as a leading-order approximation.

The resulting ranges of potentials and curvature are $\epsilon=\{6.2\times 10^{-4},0.062\}$ and $\xi=\{7.6\times 10^{-34},7.6\times 10^{-28}\}$\,cm$^{-2}$. These are indicated by the shaded grey square in Fig.~\ref{fig:exp}; we see that the PTA constraints sit slightly below-left of the SMBH points. This is commensurate with their prime target being the stochastic background of gravitational waves from unresolved sources at cosmological distances. SMBH binaries contributing to the stochastic background are likely to be at earlier phases of their inspiral than the eLISA sources\footnote{Recall that the SMBH points in Fig.~\ref{fig:parameter_space} were plotted by evaluating the gravitational fields that exist on the ISCO.}, and hence probe weaker potentials and curvatures. See \cite{Moore2014} for a useful visualization tool of gravitational wave sources and detectors. \vspace{5mm}

\noindent \textbf{Primordial Gravitational Waves.} Finally on gravitational waves, one may well ask what implications the BICEP2 results \citep{BICEP2014} have for our parameter space, if they were found to be genuinely primordial. Using equation (\ref{qem}), the Kretschman curvature around the time of inflation is so many orders of magnitude greater than anything else considered in this paper ($\sim10^{55}$~cm$^{-2}$), that we cannot sensibly mark a BICEP2 feature on our plots.

\subsection{X-Ray Timing Experiments}
A number of high-energy phenomena have been used to measure 
masses and spins of accreting neutron stars and black holes. These include quasi-periodic oscillations (QPOs) of their X-ray flux 
(\citealt{vanderKlis2000}, \citealt{Remillard2006} and references therein), 
relativistically broadened iron lines (e.g. \citealt{Miller2007}), 
as well as thermal emission from the innermost regions of the accretion disks 
\citep{McClintock2014}. All of these 
phenomena depend on, and can provide a measurement of, the locations of the 
ISCOs of compact objects, though interpreting the data requires complex modelling.

The three phenomena listed above can be used as tests of GR, provided that the quality of data is high enough to 
allow astrophysical uncertainties to be corrected for. For example, measurements of different kinds of oscillatory modes in accretion disks -- manifest as modulations of QPOs -- can allow several  multipoles of the spacetime to be disentangled \citep{Johannsen2011,Johannsen2013,Bambi2011}. 

Planned X-ray experiments such as the Large Observatory For Timing (LOFT; \citealt{Feroci2014}) and 
Athena+ (\citealt{Nandra2014}; scheduled to launch in 2028) should be able to make these measurements for black holes with masses ranging from $\sim 5 \,\mathrm{M}_\odot$ to several billions of solar masses. The yellow bounding box in the upper right of Fig.~\ref{fig:exp} marks the region probed by these X-ray experiments, where the rightmost boundary is the Schwarzchild radius (R$_S$) and the lefthand boundary marks $10$ R$_S$. Distances greater than $10$ R$_S$ become difficult to probe, because the relativistic broadening of the $\sim6.5~$keV K$\alpha$ line cannot be disentangled from other astrophysical effects on its intrinsic width. The circles themselves mark the potential and curvature at $3$ R$_S$ (the ISCO for a Schwarzchild black hole) for the same objects as in Fig.~\ref{fig:parameter_space}. Accounting for the spin of the black hole will shift these points by factors of order unity.

\subsection{The Triple System}
The recent discovery of a millisecond pulsar in the three-body system J0337+1715 \citep{Ransom2014} presents the possibility for testing gravity in the most extreme strong-field conditions to date. The system J0337+1715 consists of a millisecond pulsar and white dwarf in a near-circular binary with radius of order $3\times 10^6$~km, whilst a second white dwarf orbits the binary at a distance of $\sim 1$ AU. The large masses of all three bodies lead to significant orbital interactions and relativistic effects.

In particular, the triple system offers the opportunity for sensitive tests of the strong equivalence principle (SEP). The mechanism is identical to that described for binary pulsars in \S\ref{sub:stellar}. However, SEP-violating effects should be much stronger in the triple system because the external gravitational field, here provided by the outer orbiting white dwarf, is $\sim6-7$ orders of magnitude stronger than in the binary pulsar case.

The sensitivity of SEP tests to the structure of a body means that, arguably, the triple system J0337+1715 provides an opportunity to test gravitational fields \textit{inside} a neutron star. This extends our parameter space upwards beyond the boundaries of Fig.~\ref{fig:exp}, and closer to $\Phi=1$ than any other system. To find the exact values reached would require detailed calculations involving models for the neutron star equation of state, which are beyond the scope of the current paper. We will settle for estimating the potential and curvature at the surface of the pulsar in J0337+1715 (using the mass determined by \cite{Ransom2014} and a typical radius of $\sim 12$km -- upper pink filled circle in Fig.~\ref{fig:exp}). The nearby arrows indicate possible extension of the constraints.

\section{Conclusions}
\label{section:conclusions}

It is tempting to think that, thanks to the endeavours of the cosmological, astrophysical and laboratory communities, viable modifications of General Relativity are being squeezed into ever-tighter and more baroque corners of theory-space. The natural end result of this process, assuming that no deviations from GR+$\Lambda$CDM are detected, would be to abandon the idea of modified gravity and redirect attention towards the remaining (admittedly problematic) explanations of the dark energy problem\footnote{For example, inhomogeneous cosmologies or multiverse/landscape-related ideas.}.

The representation of a `gravity parameter space' constructed in this paper suggests that we are not yet at this stage. 
By constructing Fig.~\ref{fig:parameter_space}, the theoretical parameter space, we have found examples of unrelated phenomena that unexpectedly probe gravitational fields in the regime. For example, Big Bang Nucleosynthesis and the inspiral of binary pulsars would usually be considered as complementary tests of gravity, confirming that GR is correct both on cosmological scales at early times and on small scales at late times. In contrast, the work presented here suggests that, since they probe the same set of potentials and curvatures, combining information from BBN and pulsars does not necessarily lead to more comprehensive constraints on gravity. There is a similar pairing between the peaks of the CMB temperature power spectrum and the physics of galaxy clusters.

Furthermore, although the theoretical parameter space is densely populated, the experimental version (Fig.~\ref{fig:exp}) is much less so. Many of the curves shown there depict expectations for \textit{future} experiments rather than data available at present.

A key result of this paper has been to highlight a substantial weakness in our current constraints on gravity, namely the `curvature desert' visible in Fig.~\ref{fig:exp}. The only astrophysical systems with which we can probe this region are the interiors of galaxies; however, with our present level of understanding, the highly complex baryonic processes inside galaxies are likely degenerate with any subtle gravitational effects. The work carried out in this paper strongly motivates further investigation of precision tests of gravity on galactic and cluster scales in future, e.g. with data from the Gaia satellite.

 In Table~\ref{tab:desert} we attempt to convey a sense of the difficulty of making the required measurements, by interpreting the upper and lower boundaries of the curvature desert in a number of different physical ways. For example, the lower desert boundary of \mbox{$\xi\approx 10^{-50}$ cm$^{-2}$} is equivalent the curvature an observer would experience due to a small asteroid ($<1$~km in radius) at a distance of 1~AU in an otherwise vacuum spacetime\footnote{Perhaps a slightly more human-friendly analogy would be placing the population of planet Earth at the distance of Jupiter. Many thanks to S.~Wilkins for this observation.}. Alternatively, this is the curvature due to the sun an observer would experience from a distance of \mbox{$\sim 3\times 10^5$~AU}, roughly the distance of Alpha Centauri, our nearest star. Clearly it is very difficult to obtain clean probes of these tiny effects.
\begin{deluxetable}{ccccc}[b!]
  \tablecolumns{5}
  \tablewidth{0pt}
\tablecaption{Interpretations of the Curvature Desert}
  \tablehead{\colhead{$\xi$} & \colhead{Density}  & \colhead{$k$}  & \colhead{Mass at 1AU} & 
\colhead{Dist. from}\vspace{1mm}\\
\colhead{(cm$^{-2}$)} & (kg m$^{-3}$) & (h Mpc$^{-1}$) & (kg) & $1M_{\Sun}$ (AU) }
\startdata
$10^{-37}$ & $4.6\times 10^{-8}$ & $1.4\times 10^{6}$ & $6.5\times 10^{26}$ & $14.5$ \\
$10^{-50}$ & $4.6\times 10^{-21}$ & $0.44$ & $6.5\times 10^{13}$ & $3.1\times 10^5$ 
\enddata
\label{tab:desert}
\tablecomments{Column 2 is the density needed for a uniform sphere to have the curvature in column 1 on its surface (the radius of the sphere is not needed). Column 3 is obtained by straightforwardly interpreting $\sqrt{\xi}$ as an inverse length, with appropriate unit conversions. Columns 4 and 5 are obtained using equation (\ref{eq:xi_astro}) and solving for the appropriate quantity.}
\end{deluxetable}

It is striking that the desert sits between the stringently-tested and fully-understood regime of the Solar System, and the problem area where both dark matter and the dark energy sector come into play. If a screening mechanism shields the Solar System from non-GR effects, the curvature desert is the natural place for the transition from screened to unscreened behaviour to occur. 

Alternatively, one might argue that Fig.~\ref{fig:exp} calls into doubt the need for a screening mechanism at all. Consider a modified gravity theory in which the non-GR terms have some form of inverse dependence on $\xi$ \footnote{However, we note that theories with an inverse dependence on the Ricci scalar have been found to suffer from both theoretical and observational problems, see  \citealt{Schmidt2009}, \citealt{Sotiriou2010}, and references therein).}. With a suitable choice of functional form, the non-GR terms can be negligible in the upper half of the parameter space, yet grow to be of order unity in the cosmological regime twenty-five orders of magnitude lower in curvature. This theory would not need any explicit, non-linear screening phenomenon in order to avoid Solar System constraints -- it simply `scales' in the correct way. Likely one can construct other examples in which the scaling depends not on curvature, but on some other physical quantity. 

We stress that we are not necessarily proclaiming the belief that there exist modifications to GR which depend on the $\xi$ quantity we have made use of in this paper. The choice of axes for a gravitational parameter space is not unique, and other choices than the ones made here may be more useful for specific tasks. If one can define a mass $M$ and a distance scale $R$ that characterize a gravitational system, other physical scales (e.g. accelerations, densities, etc.) can be obtained by taking appropriate powers $p$ and $q$ in the combination $M^pR^q$. 

On our diagram (essentially a plot of $M/R^3$ versus $M/R$), these other physical scales correspond to sloped lines. Hence, if evidence for non-GR behaviour were to be found in one location of the parameter space, we would be able to draw a line(s) that divides this location from the regimes where GR works well. The slope of this line would provide an indication of the relevant quantity that controls the onset of new gravitational physics.

The fact that there are different possible `yardsticks' with which to assess gravitational fields warns us that the PPN formalism may not be sufficient for testing modern gravity theories. In essence, PPN probes only one axis direction, that of $\epsilon$, or equivalently, gravitational potential. There is need for a more sophisticated, flexible formalism which could not have been envisioned at the time PPN was developed; we leave this to a future work.

\acknowledgements

We are grateful for useful discussions with Anne Archibald, James Binney, Phil Bull, Clare Burrage, Giuseppe Congedo, Shaun Hotchkiss, Ryan Houghton, Scott Hughes, Pedro G. Ferreira, Edward Macaulay, John Miller, Scott Ransom, Jason Sanders, David Seery, Leo Stein and Steve Wilkins. TB is supported by All Souls College, Oxford. DP acknowledges support from NSF grant AST 1312034. CS acknowledges initial support from the Royal Society, and further support from the European Research Council. The research leading to these results has received funding from the European Research Council under the European Union's Seventh Framework Programme (FP7/2007-2013) / ERC Grant Agreement n. 617656 ``Theories and Models of the Dark Sector: Dark Matter, Dark Energy and Gravity''.

\appendix
\section{Gauge-Invariant Kretschmann Scalar for Cosmological Perturbations}
\label{app:gi_cosmo}
In this appendix we provide details leading to the gauge-invariant Kretschmann variable of equation (\ref{xi_cosmo}), and describe how this quantity was evaluated for the purposes of Fig.~\ref{fig:parameter_space}.

As explained in \S\ref{sub:cosmo_params}, in order to assign a curvature parameter $\xi$ to cosmological experiments we need to include linear perturbations of the Kretschmann scalar. Firstly, we set out some notation and definitions. We denote the perturbation of the Kretschmann scalar in  \textit{Fourier space} by:
\begin{align}
\label{Xidefapp}
\Xi(\vec{k},a)=K(\vec{k},a)-\0 K(a)=\Xi(k,a)\zeta(\vec{k})\;,
\end{align}
where $\0 K$ is given by equation (\ref{K0}). In the second equality we have decomposed the Fourier-space $\Xi$ into a primordial random field, $\zeta(\vec{k})$, and a $k$-dependent magnitude (which effectively acts as a transfer function for the initial perturbations). The primordial random field is the distribution of potential wells in the early universe; it is related to the primordial power spectrum $P_0(k)$ as
\begin{align}
 \langle  \zeta(\vec{k})\zeta(\vec{k}')  \rangle &= (2\pi)^3 P_0(k) \,\delta^{(3)}(\vec{k} - \vec{k}')\;.
 \label{zetadefapp}
\end{align}
For a power-law initial power spectrum with amplitude $A_s$ and spectral index $n$ we have $P_0(k)= A_s \,k^{n-4}$, where $n=1$ gives a scale-invariant dimensionless spectrum (note that the power spectra we use in this paper have dimensions of length$^{-3}$).

Now we need to find an expression for $\Xi(k,a)$ that we can evaluate. We begin by taking the linear perturbations of the coordinate-independent definition of the Kretschmann scalar, $K^2=R_{\alpha\beta\gamma\delta}R^{\alpha\beta\gamma\delta}$:
\begin{align}
K^2&=\0 R_{\alpha\beta\mu\nu} \0 R^{\alpha\beta\mu\nu}+\0 R_{\alpha\beta\mu\nu} \delta R^{\alpha\beta\mu\nu}+\0 R^{\alpha\beta\mu\nu}\delta R_{\alpha\beta\mu\nu}+{\cal O}\left(\left[\delta R_{\alpha\beta\gamma\delta}\right]^2\right)\nonumber \\
K&\approx \0 K \left[1+\frac{\0R_{\alpha\beta\mu\nu} \delta R^{\alpha\beta\mu\nu}}{\0 K^2}+\frac{\delta R_{\alpha\beta\mu\nu} \0 R^{\alpha\beta\mu\nu}}{\0 K^2}\right]^{\frac{1}{2}}\\
&=\0 K+\frac{1}{2\0 K}\left(\0R_{\alpha\beta\mu\nu} \delta R^{\alpha\beta\mu\nu}+\delta R_{\alpha\beta\mu\nu} \0 R^{\alpha\beta\mu\nu}\right)\;.
\label{qjr}
\end{align}
To evaluate the second term above, we will write the linearly perturbed line element for an FRW metric in a general gauge as
\begin{align}
\label{line_el}
ds^2&=a(\eta)^2\left[-(1+2\Psi)d\eta^2+2\nabla_iC+\left\{(1-2\Phi)\gamma_{ij}+2D_i D_jE\right\}dx^idx^j\right]\;,
\end{align}
where $\Psi,\,\Phi,\,C$ and $E$ are four scalar metric perturbations and $\gamma_{ij}$ is a flat spatial metric. The gauge-invariant Bardeen variables are related to the (non-invariant) perturbations above by
\begin{align}
\hat\Psi&=\Psi+\Hu \left(C-\dot{E}\right)+\frac{d}{d\eta}\left(C-\dot{E}\right)\\
\hat\Phi&=\Phi-\Hu\left(C-\dot{E}\right)\;.
\end{align}
We use the metric of equation (\ref{line_el}) to calculate the perturbed Riemann tensor. To evaluate equation (\ref{qjr}) we need
\begin{align}
\label{ejt}
\delta R_{\mu\nu\alpha\beta}R^{\mu\nu\alpha\beta}+ R_{\mu\nu\alpha\beta}\delta R^{\mu\nu\alpha\beta}
=&-\frac{8}{a^4}\left[6\left(\dot\Hu^2+\Hu^4\right)\hat\Psi+6\Hu^3\dot{\hat\Phi}+3\Hu\dot\Hu\left(\dot{\hat\Phi}+\dot{\hat\Psi}\right)+3\dot\Hu\ddot{\hat\Phi}+\delta^{ij}\left(\dot\Hu\partial_i\partial_j\hat\Psi-2\Hu^2\partial_i\partial_j\hat\Phi\right)\right]\nonumber\\
&-\frac{24}{a^4}\left(C-\dot{E}\right)\left[\dot\Hu\ddot\Hu+2\Hu^3\dot\Hu-2\dot\Hu^2\Hu-2\Hu^5\right]\;,
\end{align}
where we have regrouped  as much as possible of the expression into gauge-invariant variables.

Formally, we build a gauge-invariant variable by using the gauge transformation properties of any scalar perturbation $\delta\phi$ with zeroth-order value $\phi_0$: 
\begin{align}
x^\mu&\rightarrow x^\mu+\xi^\mu \quad\quad \mathrm{where}  \quad\quad \xi^\mu=\{T,\nabla_i L\} \\
\Rightarrow \quad \delta\phi&\rightarrow\delta\phi+\pounds_\xi\phi_0 \\
&=\delta\phi+\dot{\phi}_0\,T\;.
\end{align}
 Using the fact that $C$ and $E$ transform as \mbox{$C\rightarrow C-T+\dot{L}$} , \mbox{$E\rightarrow E+L$}, one of the simplest gauge-invariant variables we can construct is \mbox{$\delta\phi^{gi}=\delta\phi+\dot{\phi}_0\left(C-\dot{E}\right)$}. Unsurprisingly, using this procedure with $\phi \equiv K$ is equivalent to simply dropping the last term of equation (\ref{ejt}). 
 
Comparing equations (\ref{Xidefapp}) and (\ref{qjr}), and using the gauge-invariant construction described above, we arrive at
 \begin{align}
 \label{kretsch_phi}
 &\hat\Xi(k,a)=-\frac{1}{\0 K}\frac{4}{a^4}\Big[6\left(\dot\Hu^2+\Hu^4\right)\hat\Psi+6\Hu^3\dot{\hat\Phi}+3\Hu\dot\Hu\left(\dot{\hat\Phi}+\dot{\hat\Psi}\right)+3\dot\Hu\ddot{\hat\Phi}+\delta^{ij}\left(\dot\Hu\partial_i\partial_j\hat\Psi-2\Hu^2\partial_i\partial_j\hat\Phi\right)\Big]\;.
\end{align}
 We have used a hat to denote that the above is a gauge-invariant quantity, in line with our conventions. However, for clarity of presentation we will hereafter drop the hat. For the rest of this appendix, $\xi(k,a)$ refers to the expression in equation (\ref{kretsch_phi}).
 
We evaluate equation (\ref{kretsch_phi}) by using matter overdensities to trace potential wells. First we use the fact that $\hat\Phi=\hat\Psi$ in the late-time universe  (in GR), and then use the Poisson equation (equation~\ref{giPoisson1}) to relate $\hat\Phi$ to matter density perturbations. Because equation (\ref{kretsch_phi}) is gauge-invariant, it is trivial to convert into any gauge we choose. We will evaluate equation (\ref{kretsch_phi}) in the synchronous gauge, as this is a popular choice for Einstein-Boltzmann solver codes that calculate the evolution of cosmological perturbations (e.g. CAMB, \citealt{Lewis2000}). 

The residual gauge degree of freedom of the synchronous gauge can be used to set the velocity potential of cold dark matter to zero\footnote{Concretely, the synchronous gauge Euler equation for cold dark matter is \mbox{$\dot{\theta}_M+\Hu\theta_M=0$}, which has solution \mbox{$\theta_M\propto 1/a$}. This is the same form as the solution for the residual gauge mode of the synchronous gauge. We can choose the constant of proportionality between $\theta_M$ and the gauge mode such that they cancel; this effectively sets $\theta_M$ to zero and fully fixes the gauge. See \cite{Bucher2000,MalikWands}.}, so that we have $\Delta_M=\delta_M^{syn}$. Note that the left-hand side of equation (\ref{kretsch_phi}) does not change, because $\hat\Phi$ is gauge-invariant.

The synchronous-gauge Poisson equation is then
\begin{align}
\nabla^2{\hat\Phi}&=4\pi G_N\rho_M a^2\delta_M^{syn}\;.
\end{align}
We have dropped the summations in equation (\ref{giPoisson2}) because we are considering the late-time universe for which only dark matter and a cosmological constant are relevant, and only the dark matter clusters. Converting to Fourier space and using the Friedmann equation, we have
\begin{align}
\hat\Phi&=-\frac{3}{2}\frac{\Hu^2}{k^2}\Omega_M\delta_M^{syn}=-\frac{3}{2}\frac{H_0^2}{k^2}\frac{\Omega_{M0}}{a}\delta_M^{syn}\;.
\end{align}
We need two derivatives of this expression. We make use of the excellent approximation (in GR) that the growth rate of matter overdensities $f(a)$, defined below, behaves as $f(a)\approx\Omega^\gamma$ with \mbox{$\gamma=0.55$}  \citep{Peeblesbook, Linder2005}:
\begin{align}
\label{sfj}
f&=\frac{d\,\mathrm{ln}\Delta_M}{d\,\mathrm{ln}a}=\frac{1}{\Hu}\frac{\dot{\delta}^{syn}_M}{\delta^{syn}_M}, & \Rightarrow\quad \dot{\delta}_M^{syn}&\approx\Hu\,\delta_M^{syn}\,\Omega_M^\gamma \\
\mathrm{with}\quad\Omega_M&=\frac{\Omega_{M0}}{\Omega_{M0}+\Omega_{\Lambda 0}a^3}, &
\quad \dot{\Omega}_M&=3\Hu\Omega_M (\Omega_M-1)\label{qlk}
\end{align}
to obtain the necessary derivatives as
\begin{align}
\dot{\hat\Phi}&=\Hu\left(\Omega_M^\gamma-1\right)\hat\Phi\label{Phidot}\\
\ddot{\hat\Phi}&=E(a)\,\hat\Phi=\left[\Hu^2\left(\Omega_M^\gamma-1\right)^2+\dot\Hu(\Omega_M^\gamma-1)+3\gamma\Hu^2\Omega_M^\gamma\left(\Omega_M-1\right)\right]\,\hat\Phi\;.\label{fje}
\end{align}
The last equality above defines the function $E(a)$. We substitute equations (\ref{Phidot}) and (\ref{fje}) into equation (\ref{kretsch_phi}), obtaining the result:
\begin{align}
\label{XiLCDM}
\Xi(k,a)&=\hat\Phi [A(a)+k^2B(a)]\;,\\
\mathrm{where}\quad\quad A(a)&=-\frac{1}{\0 K(a)}\frac{4}{a^4}\Big[6\left(\dot\Hu^2+\Hu^4\right)+6\Hu^2(\Hu^2+\dot\Hu)(\Omega_M(a)^\gamma-1)+3\dot\Hu E(a)\Big]\\
B(a)&=-\frac{1}{\0 K(a)}\frac{4}{a^4}\Big[2\Hu^2-\dot\Hu\Big]\;.\label{BLCDM}
\end{align}

We need a quantity that represents an appropriate statistical average over the fluctuations present at a given time and scale. From the decomposition we made in equation (\ref{Xidefapp}), and using equation (\ref{zetadefapp}), we have
\begin{align}
\left\langle\, \Xi(\vec{k},a)\,\Xi^*(\vec{k}^\prime,a)\,\right\rangle&=  \Xi(k,a) \,\Xi^*(k^\prime,a) \langle \zeta(\vec{k}) \zeta(\vec{k}^\prime)  \rangle
\nonumber 
\\
 &=    |\Xi(k,a)|^2   (2\pi)^3 P_0(k) \,\delta^{(3)}(\vec{k} - \vec{k}')\;.\label{Xikfinal}
\end{align}
Recall that in equation (\ref{xidefinitial}) we defined our final gauge-invariant, perturbed Kretschmann quantity to plot as \mbox{$\xi^{\mathrm{cosmo}}(a)=\left[\langle|\Xi(\vec{x},a)|\rangle\right]^{1/2}$}. Hence we now need to convert back to position-space $\Xi(\vec{x},a)$. In doing so, we introduce a window function $W(k)$, as we did for the potential in \S\ref{sub:cosmo_params}, to account for the fact that a real experiment cannot access all wave numbers. Carrying out the Fourier transform, and evaluating one of the $k$ integrals using equation (\ref{Xikfinal}), we obtain
\begin{align}
\xi^{\mathrm{cosmo}}(a) = &  \sqrt{\frac{1}{2\pi^2}  \int  d\ln k \;\left[ k^3 \, |\Xi(k,a)|^2 \, P_0(k) \, W(k) \right]}\;.
\end{align}
Again we will choose the simple example window function $W(k)=\delta(\ln k-\ln k^\prime)$. This choice of window function leads to an expression for $\xi$ which is a function of scale as well as time: 
\begin{align}
\label{kab}
\xi^{\mathrm{cosmo}}(a,k) = &  |\Xi(k,a)| \sqrt{ \frac{k^3  P_0(k)}{ 2 \pi^2 } }\;.
\end{align}
The above expression is general. It is valid for any theory if $\Xi(a,k)$ is calculated directly from equation (\ref{kretsch_phi}) (which assumes only a perturbed FRW metric). In the case of $\Lambda$CDM we may use equations (\ref{XiLCDM})-(\ref{BLCDM}) instead.

For discussing the gravitational fields probed by large-scale structure, we wish to express $\xi$ in terms of the matter power spectrum, rather that the (unobservable) primordial power spectrum. To this end we use the following relations:
\begin{align}
\langle\Delta_M(\vec{k},a)\Delta_M(\vec{k}^\prime,a)\rangle&=(2\pi)^3\delta^{(3)}(\vec{k} - \vec{k}')P_M(k,a)\\
&=\left(\frac{3}{2}  \frac{H_0^2  \Omega_{0M} }{a\;k^2}\right)^{-2}\,|\hat\Phi(k,a)|^2\,(2\pi)^3\delta^{(3)}(\vec{k} - \vec{k}')P_0(k)\;.
\end{align}
The first line above is the definition of the matter power spectrum. To obtain the second line we have use the fact that $\Delta_M(\vec{k},a)$ is also defined in terms of the primordial random field and a transfer function, i.e. \mbox{$\Delta_M(\vec{k},a)=\Delta_M(k,a)\zeta(\vec{k})$}. We have also used of the usual expression for $\langle\zeta\zeta\rangle$ (equation \ref{zetadefapp}) and the Poisson equation. Evaluating the above expression at $k=k^\prime$, we obtain:
\begin{align}
\label{jad}
\left(\frac{3}{2}  \frac{H_0^2  \Omega_{0M} }{a\;k^2}\right)^2P_M(k,a)&=|\hat\Phi(k,a)|^2P_0(k)\;.
\end{align}
Finally, substituting equation (\ref{XiLCDM}) into equation (\ref{kab}) and using equation (\ref{jad}), we arrive at:
\begin{align}
\xi^{\mathrm{cosmo}}(k,a) = &\frac{3}{2}  \frac{H_0^2  \Omega_{0M} }{a}  \, \big|A(a)+k^2B(a)\big| \,  \sqrt{\frac{P_M(k,a)}{2\pi^2 k } } =   \big|A(a)+k^2B(a)\big|\,  \epsilon^{\mathrm{cosmo}}(k,a)\;.
\label{xi_cosmo_appendix}
\end{align}
This is the result we stated in equation (\ref{xi_cosmo}).

\section{Assigning $\epsilon$ and $\xi$ to the CMB Peaks}
\label{app:CMB}
\noindent We wish to find the curvature and potential probed by the peaks of the CMB. To do this we will pursue a calculation analogous to that of the CMB temperature power spectrum, ie. we will find the equivalent of the $C_\ell$s for our $\epsilon$ and $\xi$ quantities. We can then straightforwardly pick out the $\ell$-values which correspond to the CMB peaks.

We begin by transforming the gravitational potential at position $\vec{x}$ and conformal time $\eta$ to Fourier space. For a photon, the position vector $\vec{x}$ can be equivalently described by the time interval $\eta_0-\eta$ (where  $\eta_0 $ is the conformal time today) and the direction vector $\hat{n}$ that defines the lightcone of the photon. Hence: 
\begin{align}
\Phi\left(\vec{x},\eta\right)&=\Phi\left(\eta_0 - \eta, \hat{n} , \eta\right)=\int\frac{d^3k}{(2\pi)^3}\,e^{i\vec{k}\cdot\vec{x}} \,\Phi(k,\eta)\zeta(\vec{k})\;,
\end{align}
where we have decomposed the Fourier-space potential into a magnitude and the primordial random field: \mbox{$\Phi(\vec{k},\eta)=\Phi(k,\eta)\zeta(\vec{k})$}.

The potential experienced by this photon along its path is found by integrating $ \Phi\left(\eta_0 - \eta, \hat{n} , \eta\right)$ along the line-of-sight.
 As we are only treating primary anisotropies here (see \S\ref{sub:cosmo_lines}), we will introduce a visibility function that delimits the epoch of recombination and is normalized to unity, \mbox{$\int_0^{\eta_0} d\eta\, g_{\mathrm{vis}}(\eta) = 1$}.

The potential along the direction of propagation is then:
\begin{align}
\Phi\left(\hat{n}\right)&=
\int_0^{\eta_0} d\eta\,\int\frac{d^3k}{(2\pi)^3} \,e^{i\vec{k}\cdot\vec{x}}\,g_{\mathrm{vis}}(\eta)\,\Phi(k,\eta)\,\zeta(\vec{k}) \nonumber\\
&=\sum_{\ell=0}^\infty i^\ell\,\left(2\ell+1\right)\int\frac{d^3k}{(2\pi)^3} \,{\cal P}_\ell(\hat{k}\cdot\hat{n})\,\Phi_\ell(k)\,\zeta(\vec{k}) \;,\label{wer}
\end{align}
where ${\cal P}_\ell$ are the Legendre polynomials and $\hat{k}$ is a unit vector in the direction of $\vec{k}$. To reach the second line above we have used the Rayleigh formula for the expansion of the Fourier basis functions:
\begin{align}
e^{i \vec{k}\cdot\vec{x} } &=  \sum_{\ell=0}^\infty i^\ell (2\ell+1) j_\ell[k(\eta_0 - \eta)]  {\cal P}_\ell(\hat{k}\cdot\hat{n})\;,
\end{align}
where $j_\ell$ is the spherical Bessel function, and we have also defined:
\begin{align}
\Phi_\ell(k)  =  \int_0^{\eta_0} d\eta\;j_\ell[k(\eta_0 - \eta)]   \,g_{\mathrm{vis}}(\eta) \,\Phi(k,\eta)\;.
\end{align}

By analogy with the standard CMB prescription (see e.g. \citealt{Dodelson}), we decompose the field $\Phi(\hat{n})$ into spherical harmonics with coefficients:
\begin{align}
a_{\ell m}&=\int d\Omega\;Y^*_{\ell m}(\hat{n})\,\Phi(\hat{n})\\
&=\sum_{L=0}^\infty i^{L}\,\left(2L +1\right)\int\frac{d^3k}{(2\pi)^3} \Phi_L(k)\,\int d\Omega\, {\cal P}_L(\hat{k}\cdot \hat{n}) \,Y^*_{\ell m}(\hat{n})\, \zeta(\vec{k}) \;,
\end{align}
and consider the variance of these coefficients:
\begin{align}
\langle a_{\ell m}a^*_{\ell^\prime m^\prime}\rangle
& = \epsilon_\ell^2 \delta_{\ell\ell'} \delta_{m m'}\label{almdef}
\\
&=\sum_{L=0}^\infty \sum_{L^\prime=0}^\infty i^{L}(-i)^{L^\prime}\left(2L+1\right)\left(2L^\prime+1\right)\int\int\frac{d^3k}{(2\pi)^3}\frac{d^3k^\prime}{(2\pi)^3} 
 \Phi_L(k)\Phi_{L^\prime}(k^\prime)\nonumber\\
&\times \int d\Omega\;Y^*_{\ell m}(\hat{n})\,{\cal P}_{L}(\hat{k}\cdot \hat{n}) \,\int d\Omega^\prime\;Y_{\ell^\prime m^\prime}(\hat{n}^\prime){\cal P}_{L^\prime}(\hat{k}^\prime\cdot \hat{n}^\prime) \;\langle\zeta(\vec{k})\,\zeta(\vec{k}^\prime)\rangle\;.
\label{alms}
\end{align}
The angular integrals in the second line of equation (\ref{alms})
 evaluate to $\frac{4\pi }{2L+1}Y_{L m}(\hat{k}) \,\delta_{\ell L} $ and  $\frac{4\pi}{2L^\prime+1} Y^*_{L'm'}(\hat{k}^\prime)  \,\delta_{\ell' L'} $ respectively. This allows us to perform the double sum, obtaining
\begin{align}
\langle a_{\ell m}a^*_{\ell^\prime m^\prime}\rangle
&=  16 \pi^2  \; i^{\ell+\ell'}(-1)^{\ell^\prime}  \int\int\frac{d^3k}{(2\pi)^3}\frac{d^3k^\prime}{(2\pi)^3} \Phi_\ell(k)\Phi_{\ell^\prime}(k^\prime)
  Y_{\ell m}(\hat{k}) Y^*_{\ell'm'}(\hat{k}^\prime) \;  \langle\zeta(\vec{k})\,\zeta(\vec{k}^\prime)\rangle\,.
\label{alms_2}
\end{align}
We use equation (\ref{zetadefapp}) to perform one of the $k$ integrals in equation (\ref{alms_2}), and use the orthogonality relation for the spherical harmonics: \mbox{$\int d\Omega\;Y_{\ell m}(\Omega)\,Y_{\ell^\prime m^\prime}(\Omega)=\delta_{\ell\ell^\prime}\delta_{m m^\prime}$}. This leads to
\begin{align}
\langle a_{\ell m}a^*_{\ell^\prime m^\prime}\rangle
&=  \frac{2}{\pi}    \int dk \;k^2   P_0(k) |\Phi_\ell(k)|^2  \delta_{\ell \ell'} \delta_{m m'}\;,
\end{align}
where we have imposed the condition $\ell=\ell'$ in order to evaluate the coefficient $ i^{\ell+\ell'}(-1)^{\ell^\prime}$. Comparing the line above to equation (\ref{almdef}), we can read off the expression for our $\ell$-dependent measure of the gravitational potentials associated to the CMB: 
\begin{align}
\epsilon_\ell^2&=\frac{2}{\pi}\int dk\,k^2\,|\Phi_\ell(k)|^2\,P_0(k)\;.
\end{align}
The square root of the line above gives the desired expression for $\epsilon_\ell$, stated in equation (\ref{epsl}).
The calculation of $\xi_\ell$ is completely analogous to the one above, where $\Phi(k,\eta)$ is now replaced by \mbox{$\Xi(k,\eta)$}, defined in equation (\ref{Xidefapp}).

\section{Selected Data for Fig.~1 }
\label{app:data}
In Table~\ref{tab:astro} (overleaf) we present a selection of the data used in constructing in Fig.~\ref{fig:parameter_space}. The last two columns indicate the values of $\epsilon$ and $\xi$ assigned to these objects.
\hspace{-2cm}
\begin{center}
\singlespace
\begin{deluxetable}{cccccc}
  \tablecolumns{6}
\tablecaption{Stellar-Scale Astrophysical Systems}
  \tablehead{\colhead{Source} & \colhead{Mass\tablenotemark{a}}  & 
\colhead{Semi-Major Axis}  & \colhead{Eccentricity} & 
\colhead{Potential\tablenotemark{b}}, $\epsilon$ & \colhead{Curvature\tablenotemark{b}, $\xi$}\\
& \colhead{($M_\odot$)} & \colhead{(cm)} &  & & \colhead{(cm$^{-2}$)} }
\startdata
Mercury	& 1.0 & 5.79$\times 10^{12}$ & 0.206	& 3.2$\times 10^{-8}$	& 1.1$\times 10^{-32}$\\ 
Venus	& 1.0 & 1.08$\times 10^{13}$  & 0.007	& 1.4$\times 10^{-8}$	& 8.3$\times 10^{-34}$\\ 
Earth	& 1.0 & 1.50$\times 10^{13}$  & 0.0167	& 1.0$\times 10^{-8}$	& 3.2$\times 10^{-34}$\\ 
Mars	& 1.0 & 2.28$\times 10^{13}$ & 0.093	& 7.2$\times 10^{-9}$	& 1.2$\times 10^{-34}$\\ 
Jupiter	& 1.0 & 7.78$\times 10^{13}$ & 0.048	& 2.0$\times 10^{-9}$	& 2.5$\times 10^{-36}$\\ 
Saturn	& 1.0 & 1.43$\times 10^{14}$ & 0.054	& 1.1$\times 10^{-9}$	& 4.2$\times 10^{-37}$\\ 
Uranus	& 1.0 & 2.87$\times 10^{14}$ & 0.047	& 5.4$\times 10^{-10}$	& 5.0$\times 10^{-38}$\\ 
Neptune	& 1.0 & 4.50$\times 10^{14}$ & 0.009	& 3.3$\times 10^{-10}$	& 1.2$\times 10^{-38}$\\
\hline
PSR~1913$+$16   & 0.7  & 7.72$\times 10^{10}$ & 0.617 & 3.5$\times 10^{-6}$ & 2.8$\times 10^{-26}$\\ 
PSR~1534$+$12   & 0.7  & 9.21$\times 10^{10}$ & 0.274 & 1.6$\times 10^{-6}$ & 2.4$\times 10^{-27}$\\ 
Double Pulsar   & 0.65 & 3.49$\times 10^{10}$ & 0.102 & 3.0$\times 10^{-6}$ & 2.1$\times 10^{-26}$\\ 
PSR~1738$+$033  & 0.16 & 5.04$\times 10^{10}$ & 0.0   & 4.7$\times 10^{-7}$ & 1.3$\times 10^{-27}$\\ 
PSR~1012$+$5307 & 0.09 & 5.99$\times 10^{10}$ & 0.0   & 2.3$\times 10^{-7}$ & 4.5$\times 10^{-28}$\\ 
\hline
S1 & 4.31$\times 10^{6}$ & 6.33$\times 10^{16}$ & 0.496 & 2.0$\times 10^{-5}$  &	1.4$\times 10^{-37}$\\
S2 & 4.31$\times 10^{6}$ & 1.53$\times 10^{16}$ & 0.88  & 3.5$\times 10^{-4}$  &	7.2$\times 10^{-34}$\\ 
S4 & 4.31$\times 10^{6}$ & 3.71$\times 10^{16}$ & 0.406 & 2.9$\times 10^{-5}$  &	4.2$\times 10^{-37}$\\ 
S5 & 4.31$\times 10^{6}$ & 3.12$\times 10^{16}$ & 0.842 & 1.3$\times 10^{-4}$  &	3.7$\times 10^{-35}$\\ 
S6 & 4.31$\times 10^{6}$ & 5.43$\times 10^{16}$ & 0.886 & 1.0$\times 10^{-4}$  &	1.9$\times 10^{-35}$\\ 
S8 & 4.31$\times 10^{6}$ & 5.12$\times 10^{16}$ & 0.824 & 7.1$\times 10^{-5}$  &	6.1$\times 10^{-36}$\\ 
S9 & 4.31$\times 10^{6}$ & 3.65$\times 10^{16}$ & 0.825 & 1.0$\times 10^{-4}$  &	1.7$\times 10^{-35}$\\ 
S12 & 4.31$\times 10^{6}$ & 3.85$\times 10^{16}$ & 0.9   & 1.7$\times 10^{-4}$ &	7.8$\times 10^{-35}$\\ 
S13 & 4.31$\times 10^{6}$ & 3.70$\times 10^{16}$ & 0.49  & 3.4$\times 10^{-5}$ & 6.6$\times 10^{-37}$\\ 
S14 & 4.31$\times 10^{6}$ & 3.19$\times 10^{16}$ & 0.963 & 5.5$\times 10^{-4}$ &	2.7$\times 10^{-33}$\\ 
S17 & 4.31$\times 10^{6}$ & 3.88$\times 10^{16}$ & 0.364 & 2.6$\times 10^{-5}$ & 3.0$\times 10^{-37}$\\ 
S18 & 4.31$\times 10^{6}$ & 3.30$\times 10^{16}$ & 0.759 & 8.1$\times 10^{-5}$ &	8.8$\times 10^{-36}$\\ 
S19 & 4.31$\times 10^{6}$ & 9.94$\times 10^{16}$ & 0.844 & 4.2$\times 10^{-5}$ &	1.2$\times 10^{-36}$\\ 
S21 & 4.31$\times 10^{6}$ & 2.65$\times 10^{16}$ & 0.784 & 1.1$\times 10^{-4}$ &	2.4$\times 10^{-35}$\\ 
S24 & 4.31$\times 10^{6}$ & 1.32$\times 10^{17}$ & 0.933 & 7.3$\times 10^{-5}$ &	6.4$\times 10^{-36}$\\ 
S27 & 4.31$\times 10^{6}$ & 5.66$\times 10^{16}$ & 0.952 & 2.4$\times 10^{-4}$ &	2.2$\times 10^{-34}$\\ 
S29 & 4.31$\times 10^{6}$ & 4.93$\times 10^{16}$ & 0.916 & 1.6$\times 10^{-4}$ &	6.3$\times 10^{-35}$\\ 
S31 & 4.31$\times 10^{6}$ & 3.71$\times 10^{16}$ & 0.934 & 2.6$\times 10^{-4}$ &	3.05$\times 10^{-34}$\\ 
S33 & 4.31$\times 10^{6}$ & 5.11$\times 10^{16}$ & 0.731 & 4.7$\times 10^{-5}$ &	1.7$\times 10^{-36}$\\ 
S38 & 4.31$\times 10^{6}$ & 1.73$\times 10^{16}$ & 0.802 & 1.9$\times 10^{-4}$ &	1.1$\times 10^{-34}$\\ 
S66 & 4.31$\times 10^{6}$ & 1.51$\times 10^{17}$ & 0.178 & 5.2$\times 10^{-6}$ &	2.3$\times 10^{-39}$\\ 
S67 & 4.31$\times 10^{6}$ & 1.36$\times 10^{17}$ & 0.368 & 7.5$\times 10^{-6}$ &	6.9$\times 10^{-39}$\\ 
S71 & 4.31$\times 10^{6}$ & 1.32$\times 10^{17}$ & 0.844 & 3.1$\times 10^{-5}$ &	5.2$\times 10^{-37}$\\ 
S83 & 4.31$\times 10^{6}$ & 3.47$\times 10^{17}$ & 0.657 & 5.4$\times 10^{-6}$ &	2.6$\times 10^{-39}$\\ 
S87 & 4.31$\times 10^{6}$ & 1.57$\times 10^{17}$ & 0.423 & 7.1$\times 10^{-6}$ &	6.0$\times 10^{-39}$\\ 
S96 & 4.31$\times 10^{6}$ & 1.93$\times 10^{17}$ & 0.131 & 3.8$\times 10^{-6}$ &	9.5$\times 10^{-40}$\\ 
S97 & 4.31$\times 10^{6}$ & 2.72$\times 10^{17}$ & 0.302 & 3.4$\times 10^{-6}$ &	6.5$\times 10^{-40}$\\ 
SO-102& 4.31$\times 10^{6}$ & 1.24$\times 10^{16}$  & 0.68  & 1.6$\times 10^{-4}$ & 7.1$\times 10^{-35}$
\enddata
\tablenotetext{a}{The mass of the central object or, for the case of 
pulsars, the reduced mass of the binary.}
\tablenotetext{b}{The potential and curavture of the gravitational field
(see equations (\ref{eq:potential}) and (\ref{eq:xi_astro})) probed at the periastron of each orbit.}
\label{tab:astro}
\end{deluxetable}
\end{center}

\bibliographystyle{apj}

\end{document}